\newif\ifboyscout                         
\boyscouttrue 
\newif\ifhighlightedits                   
\highlighteditstrue                       
\newif\ifpreparepdf                       
\preparepdftrue 

\documentclass[3p,times,sort&compress]{elsarticle}

\usepackage{amsmath}
\usepackage{graphicx}
\usepackage{setspace}
\usepackage{pdfpages}
\usepackage{multirow}
\usepackage{tikz}
\usepackage[pdftex,colorlinks]{hyperref}
\graphicspath{{./figs/}}

\tikzset{
  treenode/.style = {align=center, inner sep=0pt, circle, black, 
    minimum size=3.5pt, draw=black},
  bn/.style = {treenode, fill=black},
  wn/.style = {treenode, fill=white},
  mytree/.style= {thick, grow=up, level distance=2mm,sibling distance=2.5mm},
}

\ifhighlightedits 
    
\else
    
\fi

\ifboyscout 
  \newcommand{\PC}[2]{$\footnotemark\footnotetext{Predrag #1: #2}$} 
  
  \newcommand{\SHK}[2]{$\footnotemark\footnotetext{SHK #1: #2}$} 
  
  \newcommand{\Xiong}[2]{$\footnotemark\footnotetext{\color{red} XD #1: #2}$} 
  
\else 
  \newcommand{\PC}[2]{}{}
  
  \newcommand{\SHK}[2]{}{}
  
  \newcommand{\Xiong}[2]{}{} 
  
\fi  

\newcommand{\iF}{integrating factor}

\newcommand{\sps}{split-step}

\newcommand{\rkip}{\RK\ methods in the interaction picture}
\newcommand{\mL}{\mathbf{L}}
\newcommand{\mN}{\mathbf{N}}
\newcommand{\RK}{Runge-Kutta}

\newcommand{\ts}{time-stepping}

\newcommand{\ats}{adaptive time-stepping}

\newcommand{\cts}{constant time-stepping}

\newcommand{\tsa}{time-step adaptation}
\newcommand{\Tsa}{Time-step adaptation}
\newcommand{\ssa}{stepsize-adaptive}
\newcommand{\Ssa}{Stepsize-adaptive}
\newcommand{\ssat}{stepsize-adaptation}

\newcommand{\sm}{slow-moving}

\newcommand{\fm}{fast-exploding}

\newcommand{\cmf}{comoving frame}

\newcommand{\stf}{static frame}
\newcommand{\fpr}{fast-phase rotation}

\newcommand{\rtol}{\texttt{rtol}}
\newcommand{\Wt}{\texttt{Wt}}
\newcommand{\Nab}{\texttt{Nab}}
\newcommand{\nab}{\texttt{nab}}
\newcommand{\Nn}{\texttt{Nn}}
\newcommand{\nn}{\texttt{nn}}
\newcommand{\Acm}{\tilde{A}}
\newcommand{\acm}{\tilde{a}}
\newcommand{\spt}{spatiotemporal}

\DeclareMathOperator{\re}{Re}

\newcommand{\cGLe}{complex Ginzburg-Landau equation}

\newcommand{\cqcGL}{cubic-quintic complex Ginzburg-Landau}

\newcommand{\cqcGLe}{cubic-quintic complex Ginzburg-Landau equation}

\newcommand{\rf}     [1] {~\cite{#1}}
\newcommand{\refref} [1] {ref.~\cite{#1}}

\newcommand{\refeq}  [1] {Eq.~(\ref{#1})}

\newcommand{\reffig} [1] {figure~\ref{#1}}

\newcommand{\refFig} [1] {Figure~\ref{#1}}

\newcommand{\reftab} [1] {table~\ref{#1}}
\newcommand{\refTab} [1] {Table~\ref{#1}}

\newcommand{\refsect}[1] {sect.~\ref{#1}}

\newcommand{\refSect}[1] {Sect.~\ref{#1}}

\newcommand{\etal}{{\em et al.}}    
\newcommand{\ie}{{i.e.}}        

\newcommand{\dmn}{-dimensional}  

\journal{Computer Physics Communications}

\begin{document}
\begin{frontmatter}

  \title{\Ssa\ integrators for dissipative solitons in \cqcGLe s}

  \author[addr1]{X. Ding\corref{cor1}}
  \ead{dingxiong203@gmail.com}

  \author[addr2]{S. H. Kang}
  \ead{kang@math.gatech.edu}

  \cortext[cor1]{Corresponding author}

  \address[addr1]{Center for Nonlinear Science, School of Physics,
    Georgia Institute of Technology, Atlanta, GA 30332-0430, USA}

  \address[addr2]{School of Mathematics,
    Georgia Institute of Technology, Atlanta, GA 30332-0430, USA}

  \begin{abstract} 
    This paper is a survey on exponential integrators to solve \cqcGLe s and related stiff problems.  
    In particular, we are interested in accurate computation near the pulsating and
    exploding soliton solutions where different time scales exist.  
    We explore \ssa\ variations of three types of exponential integrators: 
    integrating factor (IF) methods, exponential Runge-Kutta (ERK) methods 
    and split-step (SS) methods, and their embedded versions for computation and comparison.  
    We present the details, derive formulas for completeness, and consider seven different 
    \ssa\ integrating schemes to solve the \cqcGLe. Moreover, we propose using a comoving frame to 
    resolve fast phase rotation for better performance. We present thorough comparisons and
    experiments in the one- and two\dmn\  \cqcGLe s.
  \end{abstract}


\end{frontmatter}

\section{Introduction}
\label{sec:intro}
The \cGLe\rf{cross93, AKcgl02} is one of the most frequently 
studied nonlinear equations in physics and applied mathematics community. 
Derived as a general amplitude equation near the onset of instability
in the context of pattern formation\rf{cross93}, 
it has applications in various fields of physics ranging from nonlinear 
optics\rf{AkhSotTow01} to superconductivity\rf{Ginzburg04}. In order to preserve 
the invariance of the equation under a phase change 
$A\to Ae^{i\phi}$, \cGLe s only have odd-order nonlinear terms. The form with a cubic 
term is frequently used to study \spt\ chaos and intermittent traveling waves. 
Recently, \cGLe\ with an additional quintic term 
\begin{equation}
  \label{eq:cqcgl}
  A_t  = \mu A + (D_r + iD_i) \Delta A + (\beta_r + i\beta_i)|A|^2A 
  + (\gamma_r + i\gamma_i)|A|^4A 
  \,
\end{equation}
has attracted attention for its peculiar dissipative soliton solutions in 
one- and two\dmn\ 
cases\rf{Chang07, AkhSotTow01, SoAkAn00, SoAkCh01, CaCiDeBr12, Cisternas2016}. 
Here, $A(x,t)$ or $A(x, y, t)$ is a complex field. 
$\Delta$ stands for the Laplace operator $\partial_{xx}$ 
or $\partial_{xx} + \partial_{yy}$.
A dissipative soliton is a self-localized structure that arises in 
spatially extended dissipative systems. It maintains its shape temporally 
similar to a constantly propagating solitary wave packet in an optical medium.
Dissipative solitons appear as a result of the balance between dispersion
and nonlinearity and the balance between the gain and loss of energy.
Dispersion spreads the shape while nonlinearity focuses it. A nontrivial 
internal energy flow and the dependence on an external energy source differentiate 
dissipative solitons from solitons in integrable systems. Apart from a phase shift,
solitons in integrable systems remain unchanged during soliton-soliton interactions;
while dissipative solitons are free from energy or momentum conservation during
scattering and annihilation.
For a more comprehensive discussion on dissipative solitons, see \refref{Akhmediev05}.
While dissipative solitons lack most of the very special properties of
soliton solutions of integrable Hamiltonian systems, they are generic and physically 
important, both because such structures are observed for wide ranges of equation 
parameters\rf{Akhmediev95,Akhmediev95b}, and because many types of (pulsating)
soliton solutions are observed in laser experiments\rf{CuSoAk02,SoGrGrAk04,RuRuBrEr15}.

\begin{figure}[h]
  \centering
  \includegraphics[width=0.95\textwidth]{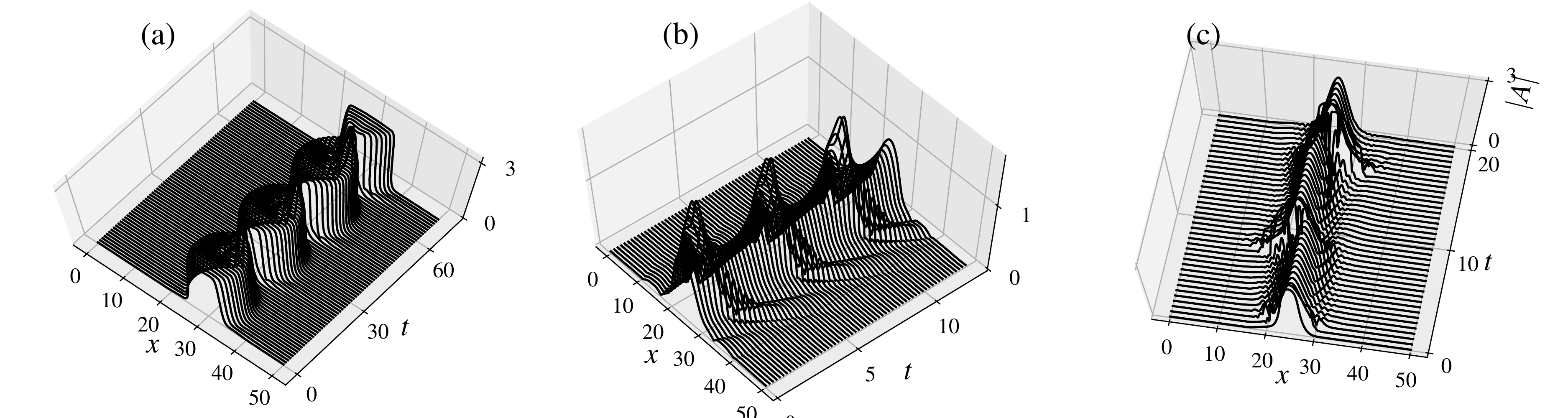}
  \caption{
    Different soliton profiles.
    The $z$-coordinate represents the magnitude of the field $|A(x,t)|$.
    In call cases, $\mu=-0.1$, $D_i=0.5$, $\gamma_r=-0.1$.
    (a) A soliton with pulsating width. 
    $D_r=0.08$, $\beta_r=0.782$, $\beta_i=1$, $\gamma_i=-0.08$.
    (b) A soliton with extremely pulsating amplitudes.
    $D_r=0.125$, $\beta_r=1$, $\beta_i=3.5$, $\gamma_i=-5$.
    (c) An exploding soliton.
    $D_r=0.125$, $\beta_r=1$, $\beta_i=0.8$, $\gamma_i=-0.6$.
  }
  \label{fig:EIDCqcgl1dSolitons}
\end{figure}

By exploring the parameter space of 
\eqref{eq:cqcgl}, Akhmediev \etal\ \rf{AkhSotTow01, SoAkAn00} identified several types of dissipative soliton solutions, such as 
plain soliton, pulsating soliton, zig-zag soliton, creeping soliton, composite soliton,
exploding soliton, chaotic soliton and so on, 
some of which are shown in \reffig{fig:EIDCqcgl1dSolitons}.
These discoveries stimulate further investigation in soliton solutions in both 
one- and two\dmn\ \cqcGLe, and the study of their bifurcations. For example,
Soto-Crespo, Akhmediev and Chiang\rf{SoAkCh01} found two coexisting solitons
with a high and a low energy respectively. 
Chang, Soto-Crespo, Vouzas and Akhmediev\rf{Chang2015b, Change15} found 
a new class of pulsating solitons with large ratios of maximal to minimal energies
as shown in \reffig{fig:EIDCqcgl1dSolitons}(b). The energy may change more than
two orders of magnitude in each period.
Tsoy and Akhmediev\rf{Tsoy2005} studied bifurcations from stationary to 
pulsating solitons based on reducing the infinite\dmn\ dynamics \eqref{eq:cqcgl}
to a five\dmn\ model. Meanwhile, Mancas and Choudhury\rf{ManCho07a} obtained a three\dmn\
model of \eqref{eq:cqcgl} by a variational method in the study of pulsating
and snake solitons. 
Among all the discoveries, solitons which undergo intermittent explosions  
stand out. As shown in \reffig{fig:EIDCqcgl1dSolitons}(c), an exploding 
soliton moves uniformly for the most time with occasional substantial changes,
after which it quickly restores back to the pre-explosion profile.
Both symmetric and asymmetric explosions\rf{Akhmediev04} are recorded and the
center of the soliton shifts after asymmetric explosions. For the two\dmn\ \cqcGLe, 
the center of an asymmetric exploding soliton exhibits a random 
walk with an anomalous diffusion rate
\rf{CaCiDeBr12, Cisternas2016}, which is unexpected for a deterministic system. 

All these novel phenomena expand our knowledge about dissipative solitons in
spatially extended chaotic systems,  and 
also propose a numerical challenge, i.e., how to efficiently integrate dissipative 
solitons. There is little concern about solitons that propagate uniformly such as
plain or composite solitons. But for pulsating or exploding solitons whose
profiles change periodically or intermittently with interleaved fast and slow
dynamics, the fast-changing parts such as the high-amplitude parts in 
\reffig{fig:EIDCqcgl1dSolitons}(b) and the
exploding parts in \reffig{fig:EIDCqcgl1dSolitons}(c) require smaller 
integration time steps. Thus \ssa\ integration schemes are preferred over
\cts\ schemes. The purpose of this paper is to introduce several \ssa\ schemes 
to integrate dissipative solitons in \cqcGLe.

For complex Ginzburg-Landau equations, finite difference 
method\rf{Wang2013}, Fourier pseudo-spectral method\rf{Taleei14} 
and Chebyshev-Tau spectral method\rf{Wang10} are applied as spatial discretization methods.
In this paper, we explore the Fourier pseudo-spectral approach, for its spectral accuracy 
when the geometry of the solution is smooth and regular.
Periodic boundary condition 
$x \in [0, L]$ or $(x, y) \in [0, L_x]\times [0, L_y]$ 
is enforced for the one- and two\dmn\ cases respectively. 
In the Fourier space, equation \eqref{eq:cqcgl} takes form of 
\begin{equation}
  \label{eq:cqcglFourier}
  \dot{a}_k = [\mu - (D_r+iD_i)q^2k^2]\,a_k + (\beta_r + i\beta_i)\,\mathcal{F}_k(|A|^2A) +
  (\gamma_r+i\gamma_i)\,\mathcal{F}_k(|A|^4A)
  \,
\end{equation}
for the one\dmn\ case, and 
\begin{equation}
  \label{eq:cqcglFourier2d}
  \dot{a}_{mn} = [\mu - (D_r+iD_i)(q_x^2m^2 + q_y^2n^2)]\,a_{mn} + (\beta_r + i\beta_i)\,
  \mathcal{F}_{mn}(|A|^2A) + (\gamma_r+i\gamma_i)\,\mathcal{F}_{mn}(|A|^4A)
  \,
\end{equation}
for the two\dmn\ case. 
Here, $q=2\pi/L$, $q_x=2\pi /L_x$ and $q_y=2\pi/L_y$.
$a_k$ and $a_{mn}$ are the $k$th Fourier mode of $A(x, t)$
and $mn$th Fourier mode of $A(x, y, t)$ respectively. 
$\mathcal{F}_k(\cdot)$ and $\mathcal{F}_{mn}(\cdot)$
denote the $k$th and $mn$th component of the one- and 
two\dmn\ discrete Fourier transform.
The linear part is represented by a diagonal matrix in the frequency domain.
The nonlinear part is first transformed back to the physical domain 
by the inverse fast Fourier transform (FFT),
integrated in the physical domain and then transformed again to the frequency domain.
This is why this strategy takes name ``pseudo-spectral''.
In soliton simulations, the size of the domain is set to
\[
  L = L_x = L_y = 50
\]
which is large enough to hold a single soliton in the domain. $1024$ and $1024^2$ Fourier 
modes are used respectively in \eqref{eq:cqcglFourier} and \eqref{eq:cqcglFourier2d}. 
The number of Fourier modes has been doubled in numerical experiments 
but without noticeable improvement in accuracy.

For the time derivative, the linear parts of \eqref{eq:cqcglFourier} and \eqref{eq:cqcglFourier2d} have quadratic 
structures and thus are stiff.  Also, to model pulsating and exploding soliton solution accurately, popular single-step integrators such as \RK\ methods requires an extremely small step size.  Therefore, we explore  exponential integrators to integrate the linear part explicitly, and use \ssat\ for exponential integrators.

In this paper,  we investigate the performance of three different types of 
\ssa\ exponential integrators. Our motivation is to tackle \cqcGLe , and to the best 
of our knowledge, this is the first work to apply these methods to dissipative soliton 
integration in \cqcGLe s. The main contributions of this paper are as follows.
First, 
We not only convert two popular integrating factor methods 
and four popular exponential Rung-Kutta methods to their \ssa\ versions, but also 
consider one \ssa\ split-step method with symmetric coefficients.  
We note that the \emph{\rkip} used in the
quantum mechanics community are equivalent to the \iF\ methods that will be
explained in \refsect{sect:if}. 
Second, 
we formulate an embedded lower-order method in a 4th-order exponential \RK\ method in 
\reftab{tab:Hochbruck_Ostermann} without an additional internal stage and are the first to 
study the embedded 5th-order exponential \RK\ method which will be introduced in
\refsect{sect:seven}.
Third, 
we utilize invariant dynamical structures in the system to accelerate the integration process.  
We compute traveling waves in \cqcGLe s and integrate the system in a \cmf\ with respect to 
the traveling waves.  This change of frame  promotes the performance of exponential \RK\ methods,
which will be covered in \refsect{sect:cm}.  
Finally, 
we experiment and compare these methods, to numerically integrate dissipative solitons 
in \cqcGLe s effectively.  

This paper is organized as follows.
In \refsect{sect:over}, we review three different types of  exponential integrators: 
integrating factor (IF) methods, exponential Rung-Kutta (ERK) methods and split-step (SS) methods.  
In \refsect{sect:seven}, we discuss how to embed a lower-order scheme in an exponential integrator
and introduce seven representative schemes. The strategy to update step size and the metrics
to evaluate the performance of an integrator is discussed in \refsect{sect:tsa}. 
\refSect{sect:exp1d} is devoted to the numerical experiments on the one\dmn\ \cqcGLe. We compare
the performance of \cts\ schemes with that of \ssa\ schemes.  In \refsect{sect:cm}, we introduce
the idea of using a \cmf\ to alleviate the \fpr\ problem.
\refSect{sect:exp2d} is devoted to the discussion of the
performance of \ssa\ schemes in the two\dmn\ \cqcGLe. We summarize our discoveries in 
\refsect{sect:concl}.

\section{Review of exponential integrators}
\label{sect:over}
Exponential integrators are formulated to solve a system of ODEs of semilinear type 
\begin{equation}
  \label{eq:sl}
  y'(t) = f(t, y) = \mL y + \mN(t, y)
  \,,
\end{equation}
which is usually spatially discretized from a parabolic PDE such as the 
\cqcGLe\ \eqref{eq:cqcgl} considered in this paper.
Here, matrix $\mL$ has a large norm or an unbounded spectrum which makes
the system stiff, and thus an ordinary explicit method is forced to use small 
time steps to ensure a stable integration process.
To cope with the stiffness, exponential
integrators treat the linear part exactly and approximate the nonlinear part 
by expansion series. There are basically three different types of single-step 
exponential integrators: \iF\ (IF) methods, exponential \RK\ (ERK) methods, and 
\sps\ (SS) methods. Both IF and ERK are \RK\ based methods, which fill out Butcher 
tables by exponential functions of $\mL$ and derive order conditions similar to
ordinary \RK\ methods. ERK methods are very popular in the applied mathematics
community. See \refref{HoOs10} for a detailed survey of IF and ERK methods.
SS methods split a single integration step into several substeps and integrate the 
system forward by considering only the linear and nonlinear effect interchangeably.
SS methods are frequently used in physics community especially in the field of nonlinear 
optics. In this section, we give an overview of IF, ERK and SS methods applied
to solving semilinear problem \eqref{eq:sl}. 
 
\subsection{IF methods}
\label{sect:if}

The integrating factor (IF) method is also called Lawson method\rf{Lawson67} which dates back to 1967.
It alleviates the stiffness of the linear part in \eqref{eq:sl} by a change of 
variables, i.e., $z(t)=e^{-t\mL}y(t)$, resulting in a nonlinear system of the new 
variable 
\begin{equation}
  \label{eq:IF}
  z'(t) = e^{-t\mL}\mN(t, e^{t\mL}z)
  \,.
\end{equation}
Here, $e^{-t\mL}$ is the integrating factor. This transformation effectively 
stabilizes this system since the linear part is integrated explicitly and the  
Jacobian of the new system
\[
  e^{-t\mL} \cdot \left.\frac{\partial N(t, y)}{\partial y}
  \right|_{y=e^{t\mL}z} \cdot e^{t\mL}
\]
has the same spectrum as that of the original one 
${\partial N(t, y)}/{\partial y}$, which is assumed to be nonstiff. 
Then popular \ts\ schemes are free to be applied to solve the new system 
\eqref{eq:IF}. Lawson\rf{Lawson67} first used the 4th-order \RK\ method to
integrate the new system, after whom, various single-/multiple-step schemes have
been implemented for \eqref{eq:IF}. (See\rf{Montanelli2016} for all these variants.)
In this paper, we apply a general $s$-stage explicit \RK\ scheme
$Y_i  = y_n + h\sum_{j=1}^{i-1} a_{ij}f(t_n + c_jh, Y_j)$ with
$y_{n+1}  = y_n + h\sum_{i=1}^{s} b_i f(t_n + c_ih, Y_i)$ to equation \eqref{eq:IF},
and obtain
\begin{equation}
  \label{eq:IF2}
  Y_i  = e^{hc_i\mL}y_n + h\sum_{j=1}^{i-1} a_{ij}e^{h\alpha_{ij}\mL} \mN(t_n + c_jh, Y_j) \,,\quad
  y_{n+1}  = e^{h\mL}y_n + h\sum_{i=1}^{s} b_i e^{h\beta_i\mL}\mN(t_n + c_ih, Y_i) \,,
\end{equation}
where $\beta_i=1-c_i$ and $\alpha_{ij}=c_i-c_j$. The local error is estimated
by embedding a lower-order \RK\ method in \eqref{eq:IF2}.
In \refSect{sect:seven}, we consider two representative \ssa\ IF schemes.

In the quantum mechanics community, IF is referred to as  
\emph{\rkip} and was first used to integrate the time-dependent Gross-Pitaevskii
equation by Caradoc-Davies\rf{Caradoc2000}. The transformation used in such case
was  $z(t)=e^{-(t-t')\mL}y(t)$ with $t'=t_n + \frac{1}{2}h_n$ during period
$t_n$ to $t_{n+1}$. Here $h_n = t_{n-1}-t_n$. The name originates form the exponential
transformation from the Schr\"odinger picture to the interaction picture and 
implies that the latter will simplify the calculation. Later, this transformation 
is applied to the (generalized) nonlinear Schr\"odinger equation by 
Johan Hult\rf{Hult07}. He compared it with SS methods and made it well
recognized in the optical fibers community. 

We note here that IF methods have a disadvantage of producing large error
coefficients when the linear term $\mL$ has a large norm, and they do not 
preserve fixed points of the original system \eqref{eq:sl}.  (For an improvement to 
generalize IF methods, see\rf{Krogstad2005}.)  However, even with these drawbacks, 
IF methods have a merit of an easy implementation, and we consider IF methods for   
\cqcGLe s.

\subsection{ERK methods}
\label{sect:erk}

Originally used in computational electrodynamics\rf{Taflove1995}, exponential Rung-Kutta (ERK) methods 
are widely used in other fields of physics\rf{Holland94, Petropoulos97, 
Schuster2000} and often referred to as  \emph{exponential time differencing} 
methods\rf{cox02jcomp, ks05com}.
An ERK method integrates the linear part exactly as IF methods. 
However, instead of using a change of variables, it resorts to an exact integration of \eqref{eq:sl}
from $t_n$ to $t_{n+1}$ by the variation-of-constants formula
\begin{equation}
  \label{eq:SLexact}
  y(t_{n+1}) = e^{h\mL}y(t_n) + h\int_0^1 e^{(1-\theta)h\mL}
  \mN(t_n+\theta h, y(t_n+\theta h)) \, d\theta
  \,.
\end{equation}
Here, $h=t_{n+1} - t_n$ is the time step. In order to approximate the integral part
above, a polynomial interpolation at $s$ non-confluent quadrature nodes 
$c_1,\cdots, c_s$ is applied to $\mN(t_n+\theta h, y(t_n+\theta h))$, \ie,
\[
  \mN(t_n+\theta h, y(t_n+\theta h)) = 
  \sum_{i=1}^s \prod_{j\neq i} \frac{\theta - c_j}{c_i - c_j} 
  \mN(t_n+c_i h, y(t_n+ c_i h))
  \,.
\]
So we get
\begin{equation}
  \label{eq:inter2}
  y(t_{n+1}) = e^{h\mL}y(t_n) + h\sum_{i=1}^s b_i(h\mL) \mN(t_n+c_i h, y(t_n+ c_i h))
  \,,\quad
  b_i(h\mL) = \int_0^1 e^{(1-\theta)h\mL}\prod_{j\neq i} \frac{\theta - c_j}{c_i - c_j} \, d\theta
  \,.
\end{equation}
Similar to an explicit \RK\ method, $y(t_n+ c_i h)$ can be approximated   
as a combination of $y(t_n + c_j h)$ for $j < i$.  Then an ERK method becomes 
\begin{equation}
  \label{eq:ETDRK}
  Y_i = e^{hc_i\mL}y_n + h\sum_{j=1}^{i-1} a_{ij}(h\mL)\mN(t_n + c_jh, Y_j) \,,\quad  
  y_{n+1} = e^{h\mL}y_n + h\sum_{i=1}^{s} b_i(h\mL)\mN(t_n + c_ih, Y_i) \,.
\end{equation}
This can also be obtained by replacing $a_{ij}e^{h\alpha_{ij}\mL}$
and $b_i e^{h\beta_i\mL}$ in \eqref{eq:IF2} with functions 
$a_{ij}(h\mL)$ and $b_i(h\mL)$. Here, coefficient functions
$a_{ij}(h\mL)$ and $b_i(h\mL)$ are chosen as linear combinations of functions 
$\varphi_j(h\mL)$, 
\[
  \varphi_j(z) =  \int_0^1 e^{(1-\theta)z}\frac{\theta^{j-1}}{(j-1)!} d\theta
  \,,\quad 
  j\ge 1
  \,.
\]
Here $b_i(h\mL)$ in \eqref{eq:inter2} is indeed a linear combination of $\varphi_j(h\mL)$. 
Furthermore, to gain a quick intuition for the choice of $\varphi_j(h\mL)$, 
we can expand the nonlinear part in \eqref{eq:SLexact} with respect to $\theta$ and get
\[
  y(t_{n+1})
  = e^{h\mL}y(t_n) + h\sum_{r=0}^{\infty} h^r \mN^{(r)}
  \int_0^1 \frac{e^{(1-\theta)h\mL}\theta^r}{r!} d\theta
  = e^{h\mL}y(t_n) + h\sum_{r=0}^{\infty} h^r \mN^{(r)}
  \varphi_{r+1}(h\mL)
  \,.
\]
Here, $\mN^{(r)}$ is the $r$th derivative of $\mN$  with respect to $\theta$. If we
define $\varphi_0(z)=e^{z}$, then $\varphi_j(z)$ has a recursion relation
\begin{equation}
  \varphi_{j+1}(z) = \frac{\varphi_{j}(z)-1/j!}{z}\,,\quad
  \varphi_{j}(0)=\lim_{z\to 0}\varphi_j(z)=\frac{1}{j!}\,,\quad
  j \ge 0 
  \label{eq:varphi}
\end{equation}
with the first few terms
\begin{equation}
  \label{eq:varphi123}
  \varphi_1(z)= \frac{e^z-1}{z}\,,\quad
  \varphi_2(z)= \frac{e^z-z-1}{z^2}\,,\quad
  \varphi_3(z)= \frac{e^z-z^2/2-z-1}{z^3} \,.
\end{equation}
  
Order conditions of ERK methods were developed
in two directions of \emph{nonstiff order conditions} and \emph{stiff order 
conditions}. Friedli\rf{Friedli1978} first derived nonstiff order conditions
up to order 5 by matching Taylor series expansions of the exact and the numerical 
solutions, which were later extended to an arbitrary order by using 
\emph{B-series}\rf{Berland05}. By expanding $a_{ij}(h\mL)$ and
$b_i(h\mL)$ in power series of $h\mL$ and truncating them to a certain order, 
nonstiff order conditions are obtained by matching the coefficients of $h$ with 
the same order on both sides of \eqref{eq:ETDRK}. 

However, this process is limiting 
when $h\mL$ has a large norm, which implies nonstiff order conditions are blind 
to the stiffness of the problem. To account for the stiffness, stiff order
conditions up to order 4 were first given in\rf{Hochbruck05}. Luan and 
Ostermann\rf{LuanOster2014_2} gave the 5th-order conditions by a perturbation 
analysis after reformulating the scheme as a perturbation of the exponential Euler
method. Later, the authors generalized the stiff order conditions to an arbitrary order by 
using \emph{exponential B-series}\rf{LuOs13}. One of the benefits of stiff order
conditions is that they preserve the equilibria of the original 
system\rf{Hochbruck05}. Stiff order conditions up to order 3 are listed in
\reftab{tab:socs}.

\begin{table}[h]
  \caption{
    Stiff order conditions for ERK up to order 3. 
    Here, $z=h\mL$.
    $J$ denotes an arbitrary square matrix.
    $\psi_{j, i}(h\mL)= \sum_{k=2}^{i-1}a_{ik}(h\mL)c_{k}^{j-1}/(j-1)! - 
    c_{i}^{j}\varphi_{j}(c_ih\mL)$.
    For the table up to order 5, see \rf{LuanOster2014_2}.
  }
  \label{tab:socs}
  \centering
  \begin{tabular}{l |l |l }
    \hline
    Order & Index & Stiff order condition \\
    \hline
    1 & 1 & $\sum_{i=1}^s b_i(z)=\varphi_1(z)$  \\
    \hline
    \multirow{2}{*}{2} & 2 & $\sum_{i=2}^s b_i(z)c_i=\varphi_2(z)$ \\
          & 3 & $\sum_{j=1}^{i-1} a_{ij}(z)=c_i\varphi_1(c_iz)\,,\quad i=2,\ldots,s$ \\
    \hline
    \multirow{2}{*}{3} & 4 & $\sum_{i=2}^s b_i(z)\frac{c_i^2}{2!}=\varphi_3(z)$ \\
          & 5 & $\sum_{i=2}^s b_i(z)J\psi_{2,i}(z)=0$ \\ 
    \hline
  \end{tabular}
\end{table}

Given a scheme with a certain nonstiff order, order reduction \rf{Hochbruck05}
may appear if this scheme has a lower stiff order, 
However, stiff order conditions are rather restrictive, and 
under favorable conditions\rf{Hochbruck05}, schemes can show a higher order of 
convergence than the order predicted by the general stiff order conditions. 
For quite a few physically interesting PDEs\rf{ks05com}, order reduction does not appear. 
The order behavior of an ERK method applied to a specific system is subtle. 
Thus, in this paper, we sort ERK methods by their nonstiff orders, and formulate \ssa\ schemes whose stiff orders match their nonstiff orders.

\subsection{SS methods}
\label{sect:ss}

The main idea of split-step (SS) methods is that if the velocity field of a physical system 
can be decomposed as a sum of several separable sub-processes, then integration 
in one step can be approximated by several consecutive substeps.  In each of the substeps 
only one sub-process takes effect. SS methods were first proposed in the 
1950s by Bagrinovskii and Godunov\rf{Bagrinovskii57}. It was also formulated by 
Strang\rf{Strang1968} as an alternating-direction difference scheme, which
has been widely used in integrating Hamiltonian systems\rf{Yoshida90} and PDEs of 
semilinear type\rf{ks05com}. 

To solve equation \eqref{eq:sl}, we split the velocity field into 
one linear and another nonlinear part.  For an $s$-stage SS method, one step 
of integration is decomposed into $2s$ substeps as follows, 
\begin{equation}
  \label{eq:SSAB}
  y_{n+1} = \phi_{\mN}(b_sh) \circ e^{a_sh \mL} \circ \cdots \circ \phi_{\mN}(b_1h) 
  \circ e^{a_1h \mL} \circ y_n
\end{equation}
here, $\circ$ is a composition operator, which means that the integration result of 
one sub-process is the input to the next sub-process. 
$\phi_{\mN}$ denotes the integration operator induced only by the nonlinear part:
\begin{equation}
  \label{eq:onlyN}
  y'(t) =  \mN(t, y)
  \,.
\end{equation}
The local error in each step can be expressed in terms of commutator 
$[\mL, \mN] = \mL\mN - \mN\mL$.
For the order condition theory of SS methods, 
see\rf{Yoshida90, Auzinger2014, Auzinger2016_2, Blanes2013}.

\section{Embedded exponential integrators}
\label{sect:seven}

In this paper, we explore time adaptive versions of numerical schemes considered in the previous sections 
for \cqcGLe s. 
Numerical schemes such as SS\rf{Mohammedi2014, CaCiDeBr12},
Adams-Bashforth\rf{Berard2015}, ERK\rf{Kassam03}, \RK\ in interaction picture\rf{Hult07} have been applied to integrate \cqcGLe s.  However, \cts\ schemes are not efficient to integrate pulsating or exploding 
soliton solutions which have different time scales as indicated in 
\reffig{fig:EIDCqcgl1dSolitons}. 
In literature, step doubling and embedded methods are the two main approaches in 
stepsize control for ordinary \RK\ methods. For exponential integrators,  
performance is a stringent concern and thus embedded 
methods are more preferable with  its less induced overhead.
Guided by this idea, 
Whalen, Brio and Moloney\rf{Whalen2015} incorporated \tsa\ into several IF and ERK schemes by 
embedding lower-order schemes. W. Auzinger and his authors\rf{Koch2013, 
Auzinger2016_2, Auzinger2016} made \tsa\ possible  in an SS method by embedding a lower- 
or same-order method.

In this section, we explore and introduce seven representative embedded schemes. 
IF4(3) and IF5(4) 
are IF based methods, where the two numbers $a(b)$ indicate that the scheme is 
$a$th-order accurate with an embedded $b$th-order scheme. ERK4(3)2(2), ERK4(3)3(3), 
ERK4(3)4(3) and ERK5(4)5(4) are ERK based methods, where four numbers $a(b)c(d)$ 
meant that the scheme has nonstiff order $a$ and stiff order $c$, and the embedded 
scheme has nonstiff order $b$ and stiff order $d$. For these six IF or ERK based 
schemes, we follow the first-same-as-last (FSAL) rule to embed lower-order schemes. 
That is, the last stage is evaluated at the same point as the first stage of the next 
step. The last scheme is SS4(3) which is based on an SS method which is 4th-order 
accurate and there is an embedded 3rd-order scheme.

\subsection{IF4(3) and IF5(4)}
\label{sect:twoIFs}

To estimate the local integration error in a general $s$-stage IF 
method described by formula \eqref{eq:IF2}, we embed a lower-order scheme of form
$ \bar{y}_{n+1}  = e^{h\mL}y_n + h\sum_{i=1}^{s+1} \bar{b}_i e^{h\beta_i\mL}
\mN(t_n + c_ih, Y_i)$. The estimated local error can be expressed as 
\[
  E_{n+1} = \bar{y}_{n+1} - y_{n+1} = h\sum_{i=1}^{s+1} (\bar{b}_i - b_i) e^{h\beta_i\mL} 
  \mN(t_n + c_jh, Y_j) \,.
\]
Here, the embedded scheme has one more stage, that the summation is from $1$ to $s+1$.
We consider two classical embedded \RK\ schemes proposed by Dormand and
Princes\rf{Dormand78, Dormand86} whose Butcher tables are tuned to minimize the 
truncation error coefficients. 
One is 4th-order accurate and the other is 5th-order
accurate, both have  one-order-lower embedded schemes. 
Their corresponding IF schemes are IF4(3) in \reftab{tab:ifrk43} and IF5(4) 
in \reftab{tab:ifrk54}. 
We note that  IF4(3) was used by Balac and Mah\'e\rf{Balac2013} in the interaction picture context.  
Following FSAL rule, in their Butcher tables, the last intermediate stage is the same as the stage of evaluating $y_{n+1}$.
The expressions for their local error estimation are listed in \reftab{tab:lte}.

\begin{table}[h]
  \caption{Butcher table of IF4(3). Here, $z=h\mL$.}
  \label{tab:ifrk43}
  \centering
  {\renewcommand{\arraystretch}{1.2}
    \begin{tabular}{c | c c c c c}
      0 & & & & & \\
      $\frac{1}{2}$  & $\frac{1}{2}e^{z/2}$ &&&&\\
      $\frac{1}{2}$ & 0 & $\frac{1}{2}$ &&&\\
      1 & 0 & 0 & $e^{z/2}$ & &\\
      1 & $\frac{1}{6}e^{z}$ & $\frac{1}{3}e^{z/2}$ & $\frac{1}{3}e^{z/2}$ & $\frac{1}{6}$ & \\
      \hline
      $b_i$ & $\frac{1}{6}e^{z}$ & $\frac{1}{3}e^{z/2}$ & $\frac{1}{3}e^{z/2}$ & $\frac{1}{6}$ & \\
      $\bar{b}_i$  &  $\frac{1}{6}e^{z}$ & $\frac{1}{3}e^{z/2}$ & $\frac{1}{3}e^{z/2}$
              & $\frac{1}{15}$  & $\frac{1}{10}$ \\
    \end{tabular}
  }
\end{table}

\begin{table}[h]
  \caption{Butcher table of IF5(4). Here, $z=h\mL$.}
  \label{tab:ifrk54}
  \centering
  {\renewcommand{\arraystretch}{1.2}
    \begin{tabular}{c | c c c c c c c}
      $0$ &&&&&&& \\
      $\frac{1}{5}$ & $\frac{1}{5}e^{z/5}$ &&&&&&\\
      $\frac{3}{10}$ & $\frac{3}{40}e^{3z/10}$ & $\frac{9}{40}e^{z/10}$ &&&&& \\
      $\frac{4}{5}$ & $\frac{44}{45}e^{4z/5}$ & $-\frac{56}{15}e^{3z/5}$ 
              & $\frac{32}{9}e^{z/2}$ &&&& \\
      $\frac{8}{9}$ & $\frac{19372}{6561}e^{8z/9}$ & $-\frac{25360}{2187}e^{31z/45}$ 
              & $\frac{64448}{6561}e^{53z/90}$ & $-\frac{212}{729}e^{4z/45}$ &&& \\
      $1$ & $\frac{9017}{3168}e^{z}$ & $-\frac{355}{33}e^{4z/5}$ & $\frac{46732}{5247}e^{7z/10}$ 
                & $\frac{49}{176}e^{z/5}$ & $-\frac{5103}{18656}e^{z/9}$ && \\
      $1$ & $\frac{35}{384}e^{z}$ & $0$ & $\frac{500}{1113}e^{7z/10}$ 
                & $\frac{125}{192}e^{z/5}$ 
                    & $-\frac{2187}{6784}e^{z/9}$ & $\frac{11}{84}$ & \\	
      \hline
      $b_i$ & $\frac{35}{384}e^{z}$ & $0$ & $\frac{500}{1113}e^{7z/10}$ 
                & $\frac{125}{192}e^{z/5}$ 
                    & $-\frac{2187}{6784}e^{z/9}$ & $\frac{11}{84}$ & \\	
      $\bar{b}_i$ & $\frac{5179}{57600}e^{z}$ & $0$ & $\frac{7571}{16695}e^{7z/10}$ 
                & $\frac{393}{640}e^{z/5}$
                    & $-\frac{92097}{339200}e^{z/9}$ & $\frac{187}{2100}$ & $\frac{1}{40}$
    \end{tabular}
  }
\end{table}

As mentioned in\rf{Whalen2015}, caution should  
be taken when ordinary embedded \RK\ methods are imported into IF methods. If 
$\alpha_{ij} < 0$ in \eqref{eq:IF2}, backward propagation in the intermediate stage
$e^{\alpha_{ij}\mL}$ is troublesome for some systems with unbounded negative linear parts.

\subsection{ERK4(3)2(2), ERK4(3)3(3), ERK4(3)4(3) and ERK5(4)5(4)}

We consider four different embedded ERK methods. The first one is 
ERK4(3)2(2)\rf{cox02jcomp} proposed by Cox and Matthews and later improved by 
Kassam and Trefethen\rf{ks05com}. This scheme has nonstiff order 4 and stiff order 2, 
and the embedded scheme has nonstiff order 3 and stiff order 2. We embed the lower-order 
scheme as shown in \reftab{tab:CoxMatthews}. Here, $\varphi_i = \varphi_i(h\mL)$ 
are the basis functions \eqref{eq:varphi} and $\varphi_{i,j}=\varphi_i(c_jh\mL)$. 
As with embedded IF methods, ERK4(3)2(2) follows the FSAL rule.

\begin{table}[h]
  \caption{
    Butcher table of ERK4(3)2(2). Here, $\varphi_i = \varphi_i(h\mL)$,
    $\varphi_{i,j}=\varphi_i(c_jh\mL)$. 
  }
  \label{tab:CoxMatthews}
  \centering
  \begin{tabular}{c | c c c c c}
    0 & & & & &\\
    $\frac{1}{2}$  & $\frac{1}{2}\varphi_{1, 2}$ &&&&\\
    $\frac{1}{2}$ & 0 & $\frac{1}{2}\varphi_{1, 3}$ &&&\\
    1 & $\frac{1}{2}\varphi_{1,3}(\varphi_{0,3}-1)$ & 0 & $\varphi_{1,3}$ & &\\
    1 & $\varphi_{1}-3\varphi_{2}+4\varphi_{3}$ & $2\varphi_{2}-4\varphi_{3}$
          & $2\varphi_{2}-4\varphi_{3}$ & $4\varphi_{3}-\varphi_{2}$ & \\
    \hline
    $b_i$  & $\varphi_{1}-3\varphi_{2}+4\varphi_{3}$ & $2\varphi_{2}-4\varphi_{3}$
          & $2\varphi_{2}-4\varphi_{3}$ & $4\varphi_{3}-\varphi_{2}$  & 0\\
    $\bar{b}_i$  & $\varphi_{1}-3\varphi_{2}+4\varphi_{3}$ & $2\varphi_{2}-4\varphi_{3}$
          & $2\varphi_{2}-4\varphi_{3}$ & 0 & $4\varphi_{3}-\varphi_{2}$ \\
  \end{tabular}
\end{table}

The second one is ERK4(3)3(3) proposed by Krogstad\rf{Krogstad2005} and is shown in 
\reftab{tab:Krogstad}. Its Butcher table is slightly different from that of ERK4(3)2(2) 
but with it gives better convergence and stability. This scheme has nonstiff order 4 and stiff 
order 3. Note, the stiff order of neither ERK4(3)2(2) and ERK4(3)3(3) matches its 
nonstiff order, but they are very popular and for moderate stiff systems. It is 
also claimed\rf{Montanelli2016} that it is hard to do much better than these two methods. 
Therefore, we consider these schemes for comparison. The embedded scheme in ERK4(3)3(3) 
is nonstiff 3rd order, stiff 3rd order. 

\begin{table}[h]
  \caption{Butcher table of ERK4(3)3(3).}
  \label{tab:Krogstad}
  \centering
  \begin{tabular}{c | c c c c c}
    0 & & & & & \\
    $\frac{1}{2}$  & $\frac{1}{2}\varphi_{1, 2}$ &&&&\\
    $\frac{1}{2}$ & $\frac{1}{2}\varphi_{1, 3} -\varphi_{2, 3}$ & $\varphi_{2, 3}$ &&&\\
    1 & $\varphi_{1,4}-2\varphi_{2,4}$ & 0 & $2\varphi_{2,4}$ & & \\
    1 & $\varphi_{1}-3\varphi_{2}+4\varphi_{3}$ & $2\varphi_{2}-4\varphi_{3}$
          & $2\varphi_{2}-4\varphi_{3}$ & $4\varphi_{3}-\varphi_{2}$ & \\
    \hline
    $b_i$ & $\varphi_{1}-3\varphi_{2}+4\varphi_{3}$ & $2\varphi_{2}-4\varphi_{3}$
          & $2\varphi_{2}-4\varphi_{3}$ & $4\varphi_{3}-\varphi_{2}$ & 0 \\
    $\bar{b}_i$  & $\varphi_{1}-3\varphi_{2}+4\varphi_{3}$ & $2\varphi_{2}-4\varphi_{3}$
          & $2\varphi_{2}-4\varphi_{3}$ & 0 & $4\varphi_{3}-\varphi_{2}$ \\
  \end{tabular}
\end{table}

The third scheme ERK4(3)4(3) is formulated by Hochbruck and Ostermann\rf{Hochbruck05},
whose Butcher table is shown in \reftab{tab:Hochbruck_Ostermann}. 
It is both nonstiff and stiff 4th-order accurate.
Since the last node coefficient $c_5$ is $1/2$ not $1$, FASL approach
fails to embed a one-order-lower scheme in this case.
Fortunately, we observe that $c_2=c_3=c_5$, so by setting
\[
  \bar{b}_2 = xb_5 \,,\quad
  \bar{b}_3 = yb_5 \,,\quad
  \bar{b}_5 = 0
\]
with other $b_i$ unchanged and choosing appropriate $x, y$, we hope to embed a 3rd-order scheme. 
\refTab{tab:socs} lists 5 stiff order conditions that should be satisfied in order to
obtain a 3rd-order scheme. The 3rd stiff order condition is already satisfied.
Setting $x+y=1$ ensures the 1st, 2nd and 4th stiff order conditions. 
Finally, setting $x=y=1/2$ make the embedded scheme satisfy a weakened but
sufficient\rf{Hochbruck05} 5th stiff order condition. The embedded scheme has
stiff order 3. We verify that it is also nonstiff order 3. 
Note, compared to the FSAL approach, this embedding does not 
require one additional internal stage; thus saves one evaluation of
the nonlinear function $\mN(t, y)$.

\begin{table}[h]
  \caption{Butcher table of ERK4(3)4(3).}
  \label{tab:Hochbruck_Ostermann}
  \centering
  \begin{tabular}{c | c c c c c}
    $0$ & & & & & \\
    $\frac{1}{2}$  & $\frac{1}{2}\varphi_{1, 2}$ &&&&\\
    $\frac{1}{2}$ & $\frac{1}{2}\varphi_{1, 3} -\varphi_{2, 3}$ & $\varphi_{2, 3}$ &&&\\
    $1$ & $\varphi_{1,4}-2\varphi_{2,4}$ & $\varphi_{2,4}$ & $\varphi_{2,4}$ && \\
    $\frac{1}{2}$ & $\frac{1}{2}\varphi_{1, 5} - 2a_{5,2}- a_{5,4}$ & $a_{5, 2}$ 
            & $a_{5,2}$ & $\frac{1}{4}\varphi_{2,5}-a_{5,2}$ &\\
    \hline
    $b_i$  & $\varphi_{1}-3\varphi_{2}+4\varphi_{3}$ & $0$
            & $0$ & $-\varphi_{2}+4\varphi_{3}$ & $4\varphi_2 - 8\varphi_3$ \\
    $\bar{b}_i$  & $\varphi_{1}-3\varphi_{2}+4\varphi_{3}$ & 
                                                             $2\varphi_2 - 4\varphi_3$
            & $2\varphi_2 - 4\varphi_3$  & $-\varphi_{2}+4\varphi_{3}$ & $0$\\
  \end{tabular}
  \[
    a_{5,2} = \frac{1}{2}\varphi_{2,5}-\varphi_{3,4}+\frac{1}{4}\varphi_{2,4}
    -\frac{1}{2}\varphi_{3,5}
  \]
\end{table}

The last scheme we consider is ERK5(4)4(4) formulated by 
Luan and Ostermann\rf{LuanOster2014} shown 
in \reftab{tab:Luan_Ostermann}. It is both nonstiff
and stiff 5th-order accurate. Following the FSAL rule, we embed a nonstiff and
stiff 4th-order scheme.  The general nonstiff order conditions are 
given in\rf{Berland05} using bi-colored trees, but conditions only up to order 4 
are listed explicitly. 
In this paper,  we derive the 5th-order conditions which is presented in \ref{sec:append}
for the completeness of the discussion. 

\begin{table}[h]
  \caption{
    Butcher table of ERK5(4)5(4). 
  }
  \label{tab:Luan_Ostermann}
  \centering
  {\renewcommand{\arraystretch}{1.1}
    \begin{tabular}{c | c c c c c c c c c}
      $0$ & & & & & & & &\\
      $\frac{1}{2}$  & $\frac{1}{2}\varphi_{1, 2}$ &&&&&&&\\
      $\frac{1}{2}$  & $\frac{1}{2}\varphi_{1, 3} -\frac{1}{2}\varphi_{2, 3}$ 
            & $\frac{1}{2}\varphi_{2, 3}$ &&&&&&\\
      $\frac{1}{4}$  & $\frac{1}{4}\varphi_{1, 4} - \frac{1}{8}\varphi_{2, 4}$
            & $0$ & $\frac{1}{8}\varphi_{2, 4}$ &&&&&\\
      $\frac{1}{2}$ & $\frac{1}{2}\varphi_{1, 5} - \frac{3}{2}\varphi_{2,5} + 
                      2\varphi_{3,5}$ & $0$ 
              & $-\frac{1}{2}\varphi_{2,5}+2\varphi_{3,5}$ 
                & $2\varphi_{2,5}-4\varphi_{3,5}$ &&&&\\
      $\frac{1}{5}$ & $\frac{1}{5}\varphi_{1, 6} - \frac{2}{25}\varphi_{2,6}
                      - \frac{1}{2}a_{64}$ & $0$ 
              & $0$ & $a_{6,4}$ & $\frac{2}{25}\varphi_{2,6}-\frac{1}{2}a_{6,4}$ \\
      $\frac{2}{3}$ & $\frac{2}{3}\varphi_{1, 7} +\frac{125}{162}a_{64}
                      - a_{75} - a_{76}$ & $0$ 
              & $0$ & $-\frac{125}{162}a_{64}$ 
                  & $a_{75}$ & $a_{76}$ \\
      $1$ & $\varphi_{1,8}-a_{85}-a_{86}-a_{87}$ & $0$ & $0$ & $0$ & $a_{85}$ 
                    & $a_{86}$ & $a_{87}$ & \\
      $1$ & $\varphi_1 - b_6-b_7-b_8$ & $0$ & $0$ & $0$ & $0$ & $b_6$ & $b_7$ & $b_8$ &\\  
      \hline
      $b_i$ & $\varphi_1 - b_6-b_7-b_8$ & $0$ & $0$ & $0$ & $0$ & $b_6$ & $b_7$ & $b_8$ & $0$\\  
      $\bar{b}_i$ & $\varphi_1 - b_6-b_7-b_8$ & $0$ & $0$ & $0$ & $0$ & $b_6$ & $b_7$ 
                        & $0$ &$b_8$\\      
    \end{tabular}
  }
  \begin{gather*}
    a_{64} = \frac{8}{25}\varphi_{2,6} - \frac{32}{125} \varphi_{3,6}\,,\quad
    a_{75} = \frac{125}{1944}a_{64} -\frac{16}{27}\varphi_{2,7} + 
    \frac{320}{81} \varphi_{3,7}\,,\quad 
    a_{76} = \frac{3125}{3888}a_{64} +\frac{100}{27}\varphi_{2,7} - 
    \frac{800}{81} \varphi_{3,7} \,, \\
    \phi = \frac{5}{32}a_{64} -\frac{1}{28}\varphi_{2,6} + 
    \frac{36}{175}\varphi_{2,7} - \frac{48}{25}\varphi_{3,7}
    +\frac{6}{175}\varphi_{4,6} + \frac{192}{35}\varphi_{4,7}
    +6\varphi_{4,8}\,,\quad
    a_{85} = \frac{208}{3}\varphi_{3,8} -\frac{16}{3}\varphi_{2,8} 
    -40\phi \,,\\
    a_{86} = -\frac{250}{3}\varphi_{3,8} + \frac{250}{21}\varphi_{2,8}
    + \frac{250}{7}\phi \,,\quad
    a_{87} = -27\varphi_{3,8} + \frac{27}{14}\varphi_{2,8} + 
    \frac{135}{7}\phi \,,\quad
    b_6 = \frac{125}{14}\varphi_{2} - \frac{625}{14}\varphi_3 
    + \frac{1125}{14}\varphi_4 \,, \\
    b_7 = -\frac{27}{14}\varphi_2 + \frac{162}{7}\varphi_{3}
    - \frac{405}{7}\varphi_4 \,,\quad
    b_8 = \frac{1}{2}\varphi_2 - \frac{13}{2}\varphi_3 
    + \frac{45}{2}\varphi_4
  \end{gather*}
\end{table}

For time adaptive method, from the Butcher table which consists of matrix functions of $h\mL$, 
it seems that the \ssa\ strategy is not efficient due to 
the cost associated with refilling the Butcher table every time the step
size $h$ is updated. 
However, for \cqcGLe s, the linear part $\mL$ is diagonal, thus evaluation of 
$\varphi_j(h\mL) $  becomes an arithmetic calculation, which has linear complexity. 
Even for systems with non-diagonal linear parts, techniques such as Krylov-subspace methods 
can be deployed to accelerate matrix function evaluation.  (See \rf{HoOs10} for more details.) 

Another implementation issue associated with ERK methods is that 
direct evaluation of $\varphi_j(z)$ in \eqref{eq:varphi123} suffers from loss of accuracy when $z$ is small. 
It is believed that the contour integral method
proposed by Kassam and Trefethen\rf{ks05com} can resolve this problem
effectively. We take this approach in our implementations. 

\subsection{SS4(3)}

We introduce one representative embedded SS scheme in this subsection.
A lower-order scheme can be embedded in \eqref{eq:SSAB} by using a 
different set of coefficients $a_i$ and $b_i$.  Unlike IF or ERK methods, 
\ssa\ SS methods do not require recalculation of any run-time coefficients, so 
\tsa\ can be implemented with nearly no additional cost. Recently, Auzinger 
and his coauthors\rf{Koch2013, Auzinger2016_2, Auzinger2016} proposed and 
optimized over 30 different embedded SS schemes with real and  complex 
coefficients $a_i$ and $b_i$. Four different strategies are suggested to estimate 
the local integration error, among which, the palindromic-pair strategy tends to 
have minimal local integration error. See\rf{Auzinger2016_2} for the details. 
Here, we focus on the palindromic-pair scheme, 
\begin{equation}
  \label{eq:SScoe}
  (a_1, b_1,\ldots, a_s, b_s) = (b_s, a_s, \ldots, b_1, a_1) 
  \,,
\end{equation}
\begin{align}
  y_1 & = \phi_{\mN}(b_sh) \circ e^{a_sh \mL} \circ \cdots \circ \phi_{\mN}(b_1h) 
  \circ e^{a_1h \mL} \circ y(t_n)  \,, \label{eq:SSy1} \\
  y_2 & = e^{b_sh \mL} \circ \phi_{\mN}(a_sh)  \circ \cdots \circ  
  e^{b_1h \mL}  \circ \phi_{\mN}(a_1h) \circ y(t_n) \,, \label{eq:SSy2} 
\end{align}
\[
  y_{n+1} = \frac{1}{2}(y_1+y_2)\,,\quad \bar{y}_{n+1} = \frac{1}{2}(y_1-y_2)
  \,.
\]
The name comes from equation \eqref{eq:SScoe} which
says that coefficients $b_i$ are totally determined by $a_i$,
\ie, $b_i = a_{s+1-i}$. States $y_1$ and $y_2$ mirror each other 
by switching roles of linear and nonlinear 
operators. They approximate $y(t_{n+1})$ with the same order 
of accuracy and their leading coefficients of the local error have 
the same magnitude but opposite signs, 
so $y_{n+1}$ is the local extrapolation of $y_1$ and $y_2$
with one more order of accuracy. $\bar{y}_{n+1}$ serves as an 
error estimator. 

In this paper, we focus on SS4(3),
a $3$-stage palindromic-pair scheme with real coefficients,
which is 4th-order accurate with an embedded 3rd-order scheme.
Auzinger called it PP3/4A in\rf{Auzinger2016}. 
The coefficients of SS4(3) are
\[
  (a_1,\; a_2,\; a_3) = (0.268330095781759925,\; 
  -0.187991618799159782,\; 0.919661523017399857)
  \,.
\]
The linear part is integrated exactly in an SS scheme \eqref{eq:SSAB}, and we solve
\eqref{eq:onlyN} $s$ times during each single step. For some systems such as the 
cubic \cGLe\ and the nonlinear Schr\"odinger equation, \eqref{eq:onlyN} can be solved
explicitly\rf{Wang2013}, but such explicit formula does not exist for the \cqcGLe.
We use numerical schemes to solve \eqref{eq:onlyN}, in particular, the 4th-order \RK\ scheme.   
Since there are many evaluations of the nonlinear function $\mN(t, y)$ in a single step 
in \eqref{eq:SSy1} and \eqref{eq:SSy2}, we restrict ourselves to only considering SS4(3) 
in this paper.

\section{\Tsa\ and performance metrics}
\label{sect:tsa}

Each \ssa\ scheme mentioned in \refsect{sect:seven} consists of a higher-order scheme and 
an embedded lower-order scheme. Though the local error is estimated for the lower-order 
scheme, following the local extrapolation strategy, we take the higher-order scheme to 
integrate the system forward. 
Integration accuracy is locally maintained by rejecting the current step size if the 
estimated local error is larger than the specified tolerance or accepting the state 
calculated by the current step size if otherwise. When the step size is rejected,
a smaller step size is chosen to recompute the next state. If the calculated
state is accepted then step size is scaled up for future computation.
Let \rtol\ be the relative tolerance for the local error, and 
we maintain 
\[
  ||E_{n+1}||_\infty < \rtol\cdot ||y_{n+1}||_\infty
\]
at each integration step.
Here, $E_{n+1}$ is the estimated local error for $y_{n+1}$. See expressions of $E_{n+1}$ 
for all seven schemes in \reftab{tab:lte}. 
Then the attempted new step size is
\[
  h_{attempt} = s\cdot h\,,\quad 
  s = v\left(\frac{\rtol \cdot ||y_{n+1}||_\infty}
    {||E_{n+1}||_\infty} \right)^{1/p}
  \,.
\]
$p$ is the order of the local error of the 
embedded scheme. $v$ is a safe factor and is set to $0.9$. Updating step size for each 
single step is not efficient because of the frequent recalculation of $h\mL$ and other 
dependent coefficients. Also, to avoid step size oscillation, we update step size only 
when $s < 1$ or when the difference between $h_{attempt}$ and $h$ is large enough. So we 
adopt the \emph{lazy adaptation} strategy as in \rf{Whalen2015}:
\[
  \mu = 
  \begin{cases}
    0.4   & \quad  s < 0.4 \\
    s     & \quad  0.4 \le s < 0.85 \\
    0.85  & \quad  0.85 \le s < 1 \\
    1     & \quad  1 \le s < 1.25 \\
    s     & \quad  1.25 \le s < 4 \\
    4     & \quad  s \ge 4
  \end{cases}
\]
The rule for updating step size is then $h_{new} = \mu h$.

In order to compare the performance of different schemes, 
the following metrics are used.
\begin{itemize}
\item \Nn: Number of evaluations of nonlinear function $\mN(t, y)$ during
  the whole integration process. 
  For complex systems, evaluations of $\mN(t, y)$ 
  take the majority part of the total integration time, thus we use 
  \Nn\ as one of the main metrics to compare different methods.
  \nn\ denotes the number of evaluations of $\mN(t,y)$ in a single step, 
  whose values for seven different schemes are listed in \reftab{tab:lte}.
  Note that, in SS4(3), we use the 4th-order \RK\ scheme to
  solve the nonlinear propagation equation \eqref{eq:onlyN}. Thus its 
  \nn\  entry is 24.
  
\item \Nab: Number of calculations of coefficients $a_{ij}$ or $b_i$ during
  the whole integration process. \nab\ denotes the number of distinct 
  $a_{ij}$ and $b_i$ entries in a Butcher table.
  Elements in Butcher tables of IF and ERK methods are exponential functions 
  like $e^{h\mL}$, which need to be recalculated whenever the step 
  size is updated. Moreover, coefficients $a_{ij}$ and $b_{i}$ in ERK methods 
  are evaluated by contour integrals, which need more time to calculate than 
  those in IF methods.  Thus we only consider evaluations of $a_{ij}$ and 
  $b_{i}$ in ERK methods. \refTab{tab:lte} lists the \nab\ values of 
  four ERK schemes. Note, \nab\ of ERK4(3)2(2) is 4 not 6 thanks to an
  implementation strategy from\rf{ks05com}. 

\item Global relative accuracy:  By using a very small step size, one can obtain a
  solution relatively close to the ``true'' solution.    
  The global relative accuracy of each \ssa\ scheme is then calculated.
  
\item \Wt : wall-clock time used for the integration, which is measured on a desktop 
  equipped with 6 Intel i7-4930K 3.40GHz cores and 32G memory.
  
\end{itemize}

\begin{table}[h]
  \caption{
    Characteristics of the  seven embedded schemes. 
    `order': the order of the local error of the embedded scheme.
    `\nn': the number of evaluations of $\mN(t,y)$ in a single step.
    `\nab': the number of distinct $a_{ij}$ and $b_i$ entries in a Butcher table.
    `$E_{n+1}$': the expression for the estimated local error. 
  }
  \label{tab:lte}
  \centering
  {\renewcommand{\arraystretch}{1.4} 
    \begin{tabular}{l | c | c | c | c }
      \hline
      scheme & order & \nn & \nab & $E_{n+1}$\\
      \hline
      IF4(3) & 4 & 5 & - & $\frac{1}{10}h\left[\mN(t_n+h, Y_5)- \mN(t_n+h, Y_4)\right]$ \\
      \hline
      \multirow{3}{*}{IF5(4)} &  \multirow{3}{*}{5} & \multirow{3}{*}{7}
                           & \multirow{3}{*}{-}
                                  & $h\left\{\frac{-71}{57600}\mN(t_n, Y_1) + 
                                    \frac{71}{16695}\mN(t_n+\frac{3h}{10}, Y_3) 
                                    \right.$ \\
             & &&& $-\frac{71}{1920}\mN(t_n+\frac{4h}{5}, Y_4)
                   +  \frac{17253}{339200}\mN(t_n+\frac{8h}{9}, Y_4) $ \\
             &&&&  $\left. -\frac{22}{525}\mN(t_n+h, Y_5) +  \frac{1}{40}\mN(t_n+h, Y_6) \right\}$ \\ 
      \hline     
      ERK4(3)2(2) & 4 & 5 & 4 & $ b_4h\left[\mN(t_n+h, Y_5)-
                                \mN(t_n+h, Y_4)\right]$ \\
      \hline
      ERK4(3)3(3) & 4 & 5 & 8 & $b_4h\left[\mN(t_n+h, Y_5)-
                                \mN(t_n+h, Y_4)\right]$ \\
      \hline
      ERK4(3)4(3) & 4 & 5 & 11 &  $b_5h\left[\mN(t_n+h, Y_5)-
                                 \frac{1}{2}(\mN(t_n+h/2, Y_2)+\mN(t_n+h/2, Y_3))\right]$ \\
      \hline
      ERK5(4)5(4) & 5 & 9 & 23 & $b_8h\left[\mN(t_n+h, Y_9)-
                                 \mN(t_n+h, Y_8)\right]$ \\
      \hline
      SS4(3) & 4 & 24 & - & $(y_1-y_2)/2$ \\
      \hline
    \end{tabular}
  }
\end{table}

\section{Numerical experiments and comparisons}
\label{sect:exp1d}

\begin{figure}[h]
  \centering
  \includegraphics[width=0.95\textwidth]{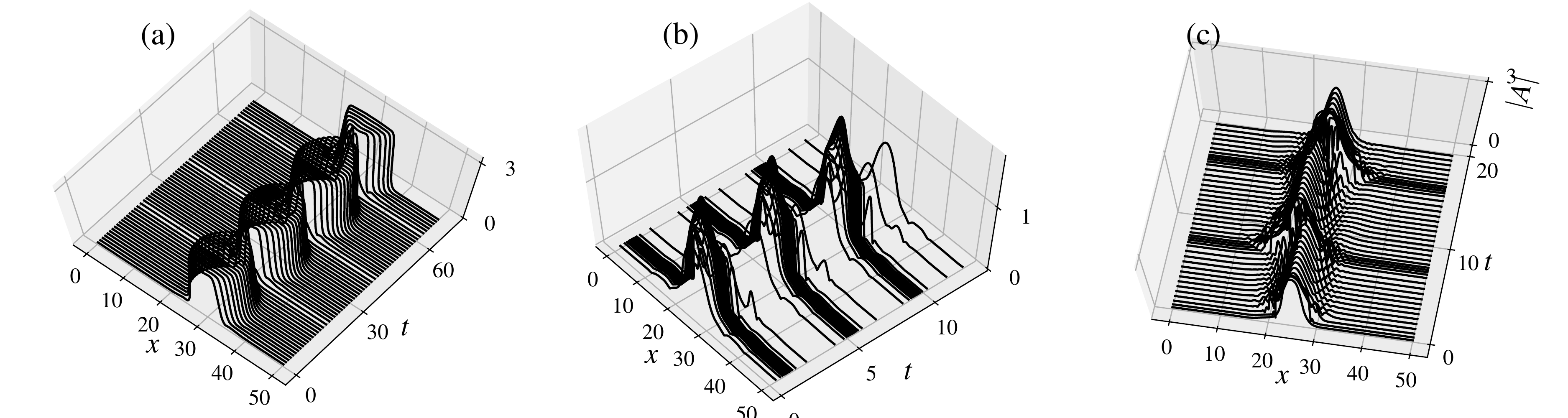}
  \caption{
    Integration of solitons in \reffig{fig:EIDCqcgl1dSolitons}
    by ERK4(3)2(2) with relative tolerance $\rtol = 10^{-10}$.
    The spacing between consecutive profiles indicates the relative 
    magnitude of integration step size.
  }
  \label{fig:EIDCqcgl1dSolitonsAdapt}
\end{figure}

We first show the performance of ERK4(3)2(2) for the three different 
soliton solutions displayed in \refFig{fig:EIDCqcgl1dSolitons}, and then 
compare the performance of the seven \ssa\ schemes specifically for the 
exploding soliton solution. 
\refFig{fig:EIDCqcgl1dSolitonsAdapt} shows the integration results of ERK4(3)2(2).
The spacing in these fence plots indicates the relative magnitudes of the step
sizes. The pulsating soliton in panel (a) is integrated almost 
with a constant step size as in \refFig{fig:EIDCqcgl1dSolitons}(a), which is anticipated
since there is only one time scale in the dynamics of this soliton.
On the other hand,  \tsa\ slows down dramatically the integration of the high-spike parts for 
the extremely pulsating soliton in panel (b), and the performance of other 
six \ssa\ schemes is similar for this soliton. This observation indicates
that \ssa\ methods efficiently
integrate extremely pulsating solitons.  For the exploding soliton in panel (c), 
ERK4(3)2(2) slows down the exploding parts moderately.

\begin{figure}[h]
  \centering
  \begin{minipage}{.26\textwidth}
    \centering \small{\texttt{(a)}}
    \includegraphics[width=\textwidth]{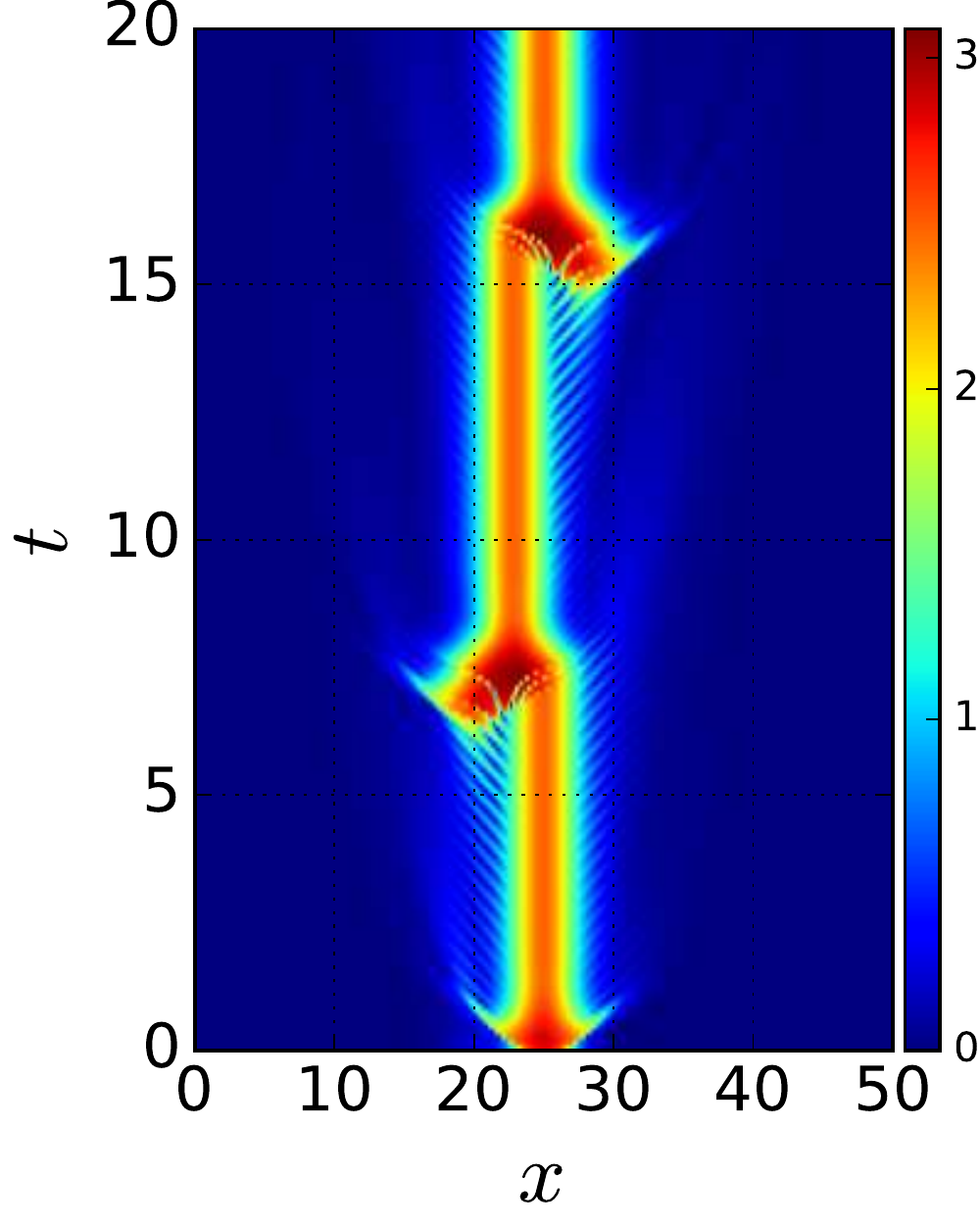}
  \end{minipage}
  \begin{minipage}{.26\textwidth}
    \centering \small{\texttt{(b)}}
    \includegraphics[width=\textwidth]{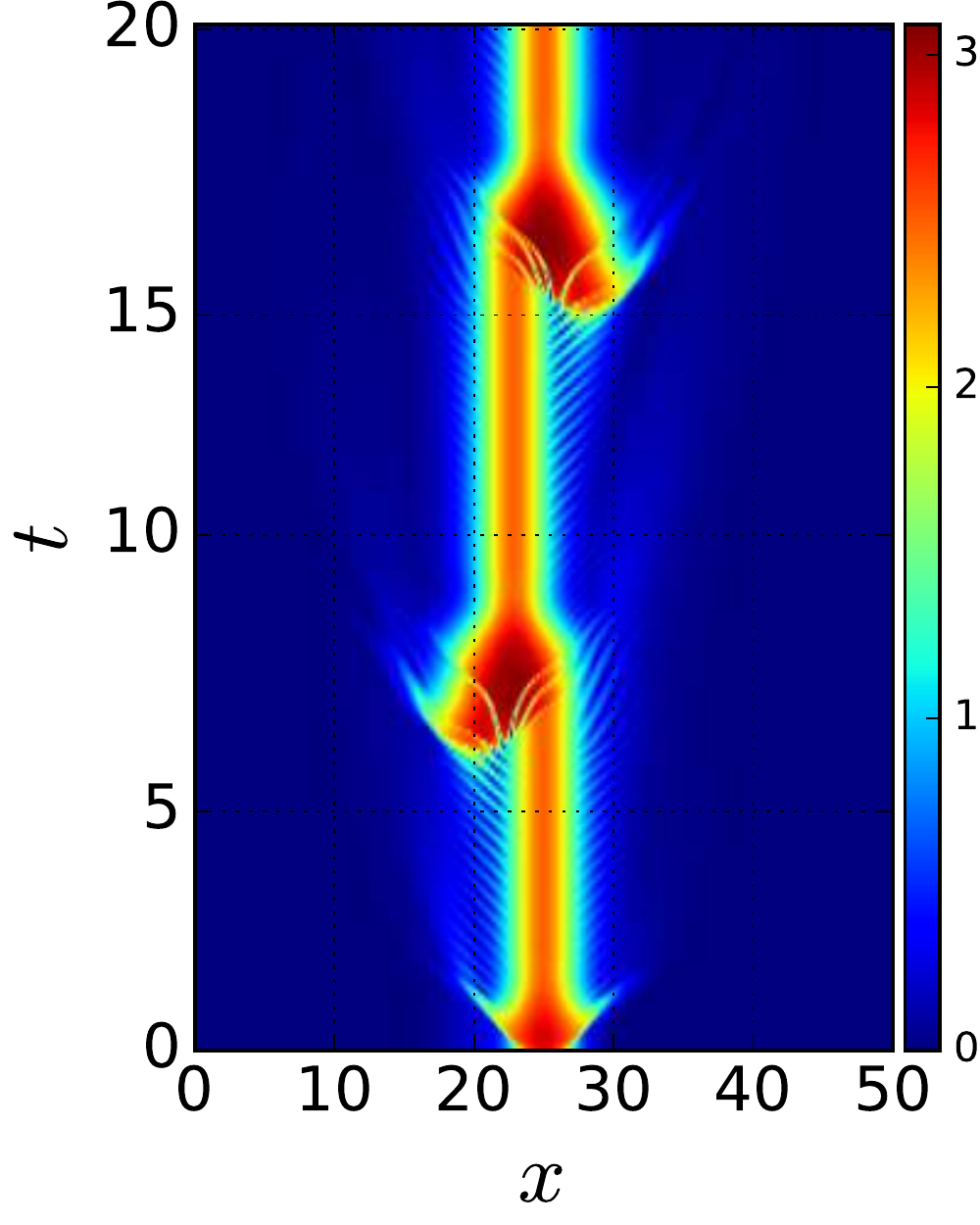}
  \end{minipage}
  \begin{minipage}{.26\textwidth}
    \centering \small{\texttt{(c)}}
    \includegraphics[width=\textwidth]{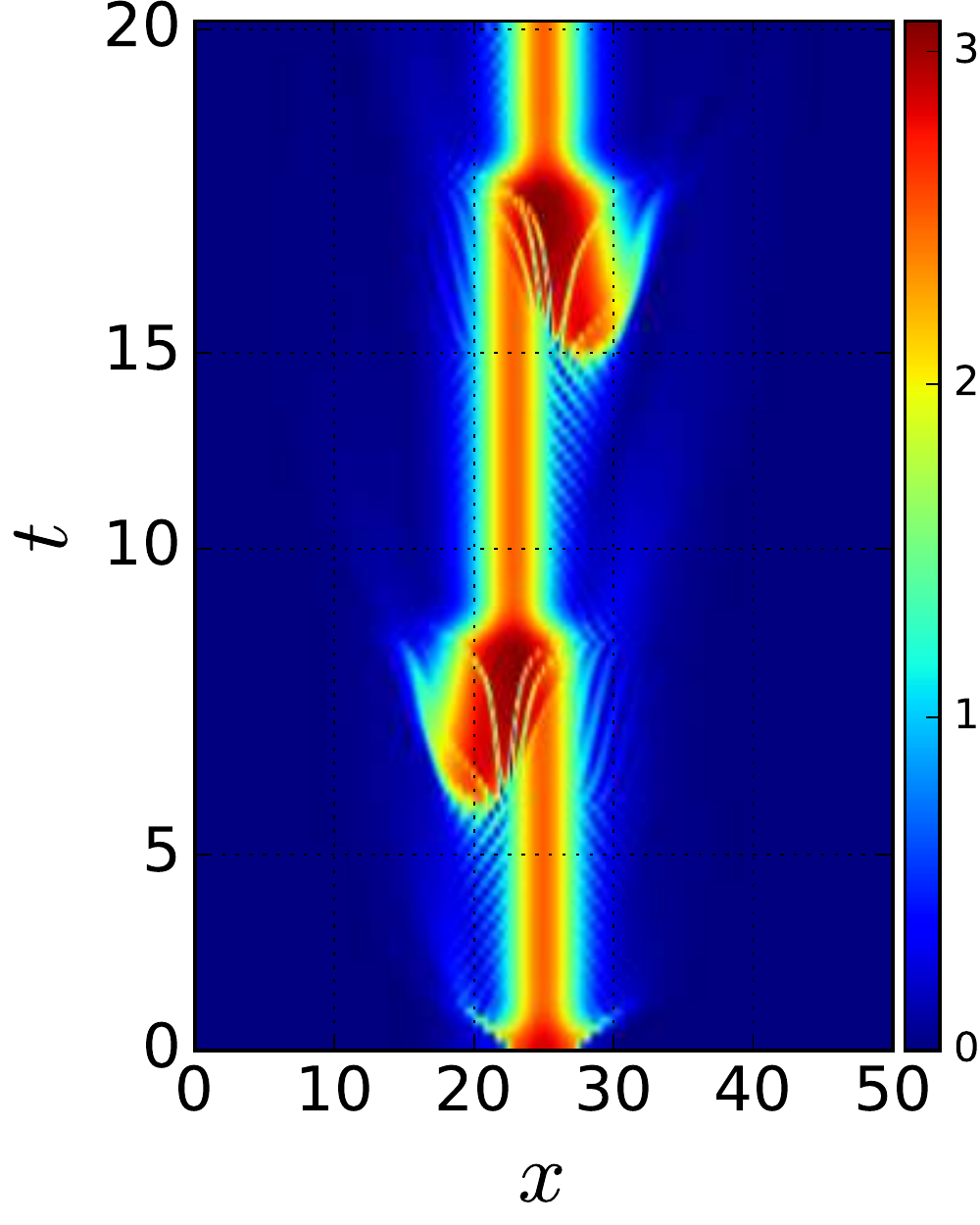}
  \end{minipage}
  \caption{
    Heat maps of integration of initial condition \eqref{eq:init} for 
    $t\in[0, 20]$. The color represents the magnitude of 
    $|A(x,t)|$.
    (a) \cts\ method with step size  $h=2\cdot 10^{-3}$. 
    (b) ERK4(3)2(2) and (c) SS4(3) both with relative 
    tolerance $\rtol = 10^{-10}$. 
  }
  \label{fig:EIDCqcgl1d_heat}
\end{figure}

\begin{figure}[h]
  \centering
  \begin{minipage}{.47\textwidth}
    \centering \small{\texttt{(a)}}
    \includegraphics[width=\textwidth]{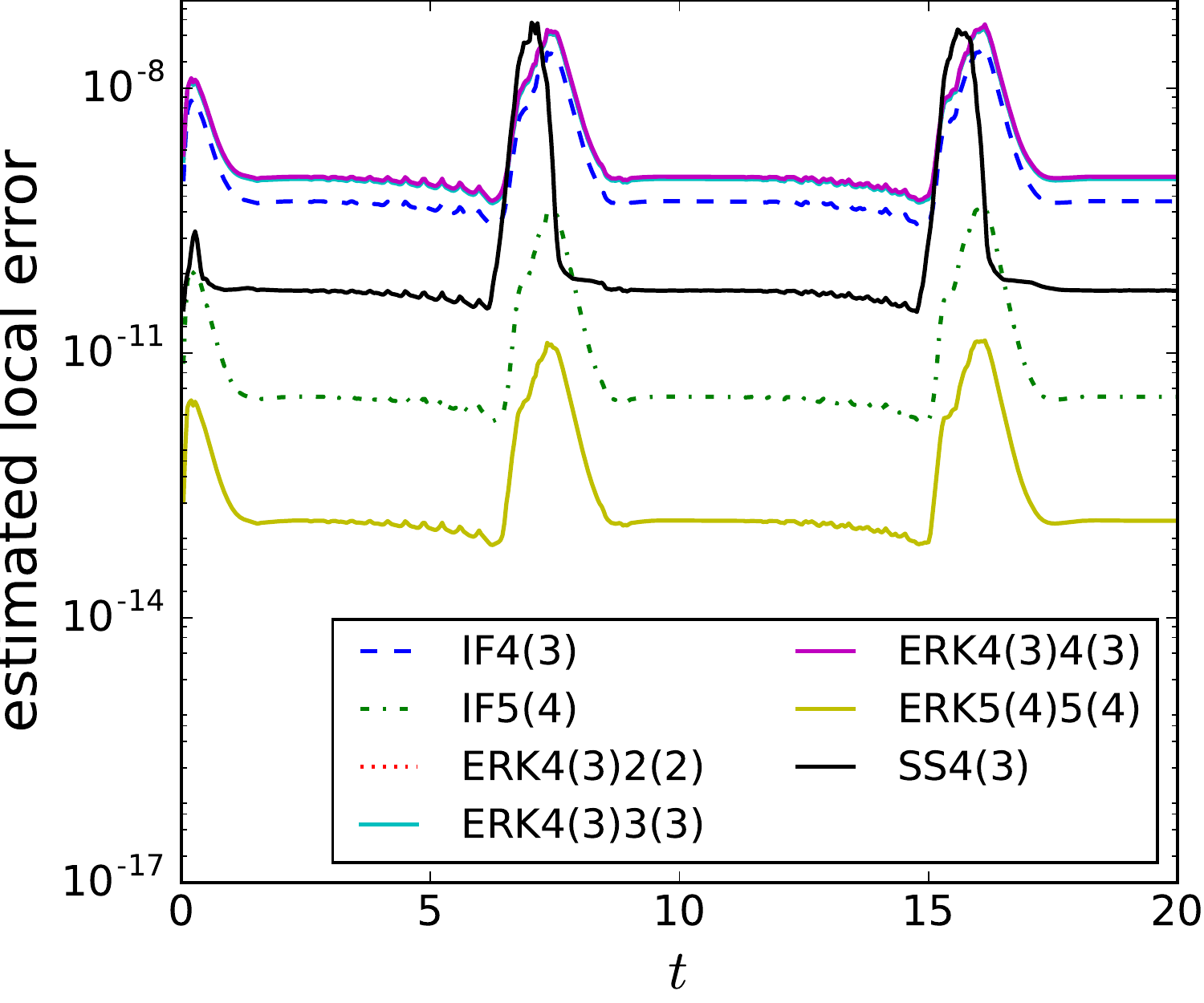}
  \end{minipage}
  \begin{minipage}{.47\textwidth}
    \centering \small{\texttt{(b)}}
    \includegraphics[width=\textwidth]{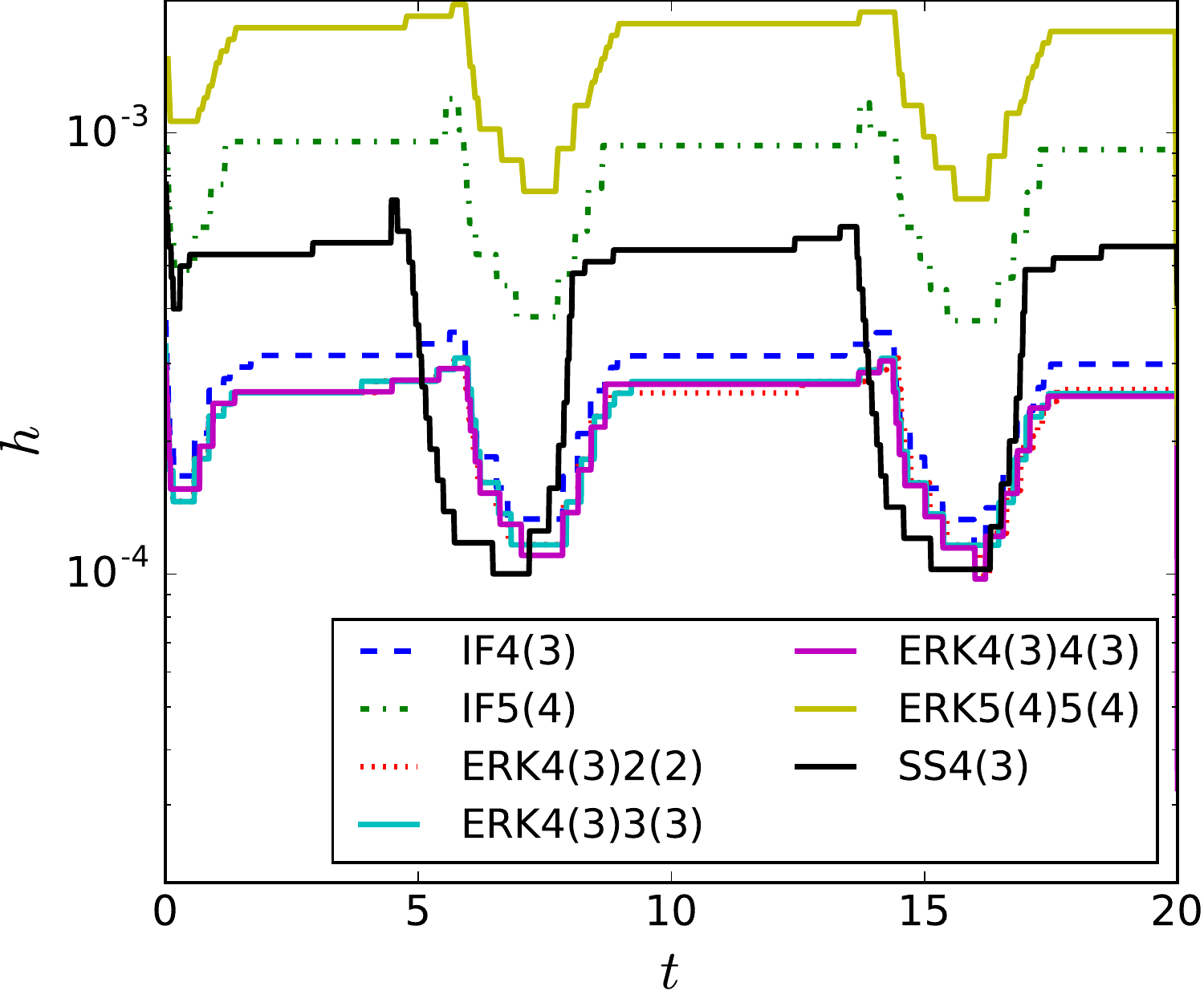}
  \end{minipage}
  \caption{
    (a) Estimated local errors for the \cts\ result in \refFig{fig:EIDCqcgl1d_heat}(a).
    (b) Step sizes used during $t\in[0, 20]$ for ERK4(3)2(2) in 
    \refFig{fig:EIDCqcgl1d_heat}(b).
  }
  \label{fig:EIDCqcgl1d_static_h}
\end{figure}

The initial condition that generates the exploding soliton in 
\refFig{fig:EIDCqcgl1dSolitonsAdapt}(c) is 
\begin{equation}
  \label{eq:init}
  A(x, 0) \,=\, 2.5\exp\left(
    -450\left(\frac{x}{L} - \frac{1}{2} \right)^2 
  \right) + 
  0.2\exp\left(
    -450\left(\frac{x}{L} - \frac{2}{5} \right)^2
  \right)
  \,,
\end{equation}
which is a Gaussian wave in the middle of the domain 
composed with a small perturbation on the left side. 
\refFig{fig:EIDCqcgl1d_heat} shows the integration result in the
form of heat map for the \cts\ method, ERK4(3)2(2) and SS4(3) respectively.
Different from \refFig{fig:EIDCqcgl1dSolitonsAdapt}(c), the heat maps
scale the time axis (y-axis) to indicate the magnitude of step sizes.
Two asymmetric explosions appear during the 
integration time window $t\in[0, 20]$. 
The explosions take small fractions of the  total integration time. The soliton profile does not change for the rest. 
\refFig{fig:EIDCqcgl1d_static_h} shows the estimated local integration error 
for \refFig{fig:EIDCqcgl1d_heat}(a) and 
the step sizes used by ERK4(3)2(2) in \refFig{fig:EIDCqcgl1d_heat}(b).
During the \fm\ parts, the estimated local error bursts substantially, spanning 2 to 3 
orders of magnitude. The \fm\ parts are the main cause of the accuracy lose. 
Step size is reduced when 
explosions happen and return to the normal level after explosions end. 
As shown in \refFig{fig:EIDCqcgl1d_static_h}(b), IF4(3), 
ERK4(3)2(2), ERK4(3)3(3) and ERK4(3)4(3) have almost the same adaptation 
pattern, while SS4(3) behaves more aggressively in the \sm\ parts. This is the reason for the 
explosions in \refFig{fig:EIDCqcgl1d_heat}(c) are more stretched than those in 
\refFig{fig:EIDCqcgl1d_heat}(b). Moreover, in 
\refFig{fig:EIDCqcgl1d_static_h}(b) the two holes of SS4(3) are slightly shifted to the 
left side compared to other 4th-order schemes, because SS4(3) uses large step 
sizes during \sm\ parts. 
\begin{figure}[!ht]
  \centering
  \includegraphics[width=0.9\textwidth]{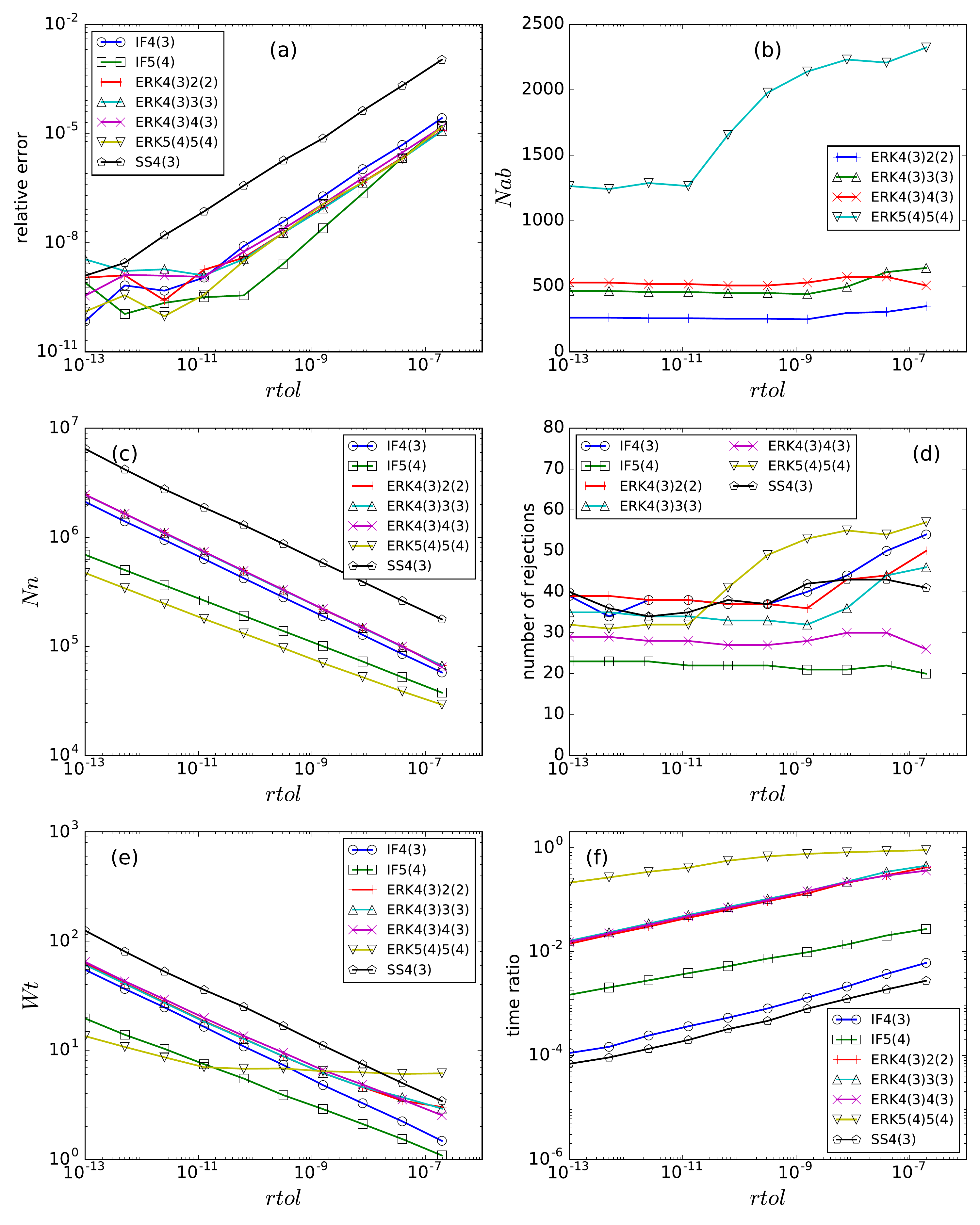}
  \caption{
    The performance of the seven \ssa\ schemes for integrating the 
    exploding soliton with initial condition \refeq{eq:init}.
  }
  \label{fig:EIDCqcgl1d_static}
\end{figure}
In \refFig{fig:EIDCqcgl1d_heat}, the \fm\ parts in panel (b) and (c) are stretched 
slightly compared with panel (a) and SS4(3) is than ERK4(3)2(2).  We further explore this in \refsect{sect:cm}. 

We compare the performance of the seven \ssa\ schemes.
\refFig{fig:EIDCqcgl1d_static} shows the performance of all the seven schemes 
for integrating the exploding soliton.
Panel (a) shows that \rtol\ effectively controls the relative global error of all 
schemes. A smaller \rtol\ usually produces a more accurate result. However, such tendency 
saturates for $\rtol < 10^{-11}$. SS4(3) has the largest relative global error since 
SS4(3) uses larger step sizes during \sm\ parts as shown in 
\refFig{fig:EIDCqcgl1d_static_h}(b). The integrator should spend more time on the \fm\ 
parts, and with an appropriate \rtol\ one can achieve a required global accuracy. Panel 
(c) shows the number of evaluations of the nonlinear function \Nn\ versus \rtol. The two 
5th-order schemes IF5(4) and ERK5(4)5(4) have the least \Nn\ because they use far fewer 
steps even though there are more evaluations of the nonlinear function in each step. 
Except for SS4(3), all other 4th-order schemes share a similar behavior of \Nn. 
SS4(3) has a much larger \Nn\ because there are 24 evaluations of $\mN(t, y)$ in a single 
step for SS4(3). Panel (e) plots the wall time elapsed in seconds versus \rtol. IF4(3), 
IF5(4) and SS4(3) have the same tendency as in panel (c). However, the relation saturates 
for ERK methods for a large \rtol, which is most significant for ERK5(4)5(4). 
For an ERK method, the time used to refilling its Butcher table 
takes a larger percentage when \rtol\ increases as shown in panel (f). 

For ERK methods, refilling a Butcher table involves
recalculation of $a_{ij}$ and $b_j$ which constitutes a large part of the total time.
As shown in panel (b), \Nab\ of ERK5(4)5(4) increases 
substantially when \rtol\ gets beyond $10^{-11}$, and for ERK4(3)2(2),
ERK4(3)3(3) and ERK4(3)4(3), \Nab\ becomes slightly larger when \rtol\ increases
to $10^{-7}$. The time used for refilling Butcher tables dominates \Wt\
at large \rtol\ for ERK methods. This phenomenon raises a question: 
why does \Nab\ increase when \rtol\ becomes large enough? \Nab\ is proportional to 
the number of times that a step size is rejected during the integration process. 
When \rtol\ becomes larger, the step size is larger and thus there is an 
increased possibility of step-size oscillation in the integration process. Panel 
(d) shows that the number of rejections increases as \rtol\ increases. The percentage of 
the time used to recalculate $h\mL$-dependent coefficients  increases, as shown in panel (f). For ERK methods, 
the time spent on refilling the Butcher table of ERK5(4)5(4) almost takes the whole 
computation time when \rtol\ reaches $10^{-9}$ as shown in panel (f).

The \ssa\ schemes effectively control the local
error during the integration process. The two 5th-order schemes IF5(4) and 
ERK5(4)5(4) have the best performance for exploding solitons, 
but the performance of ERK5(4)5(4) deteriorates 
when \rtol\ becomes too large.

\section{Comoving-frame improvement for ERK methods}
\label{sect:cm}
\begin{figure}[h]
  \centering
  \includegraphics[width=0.45\textwidth]{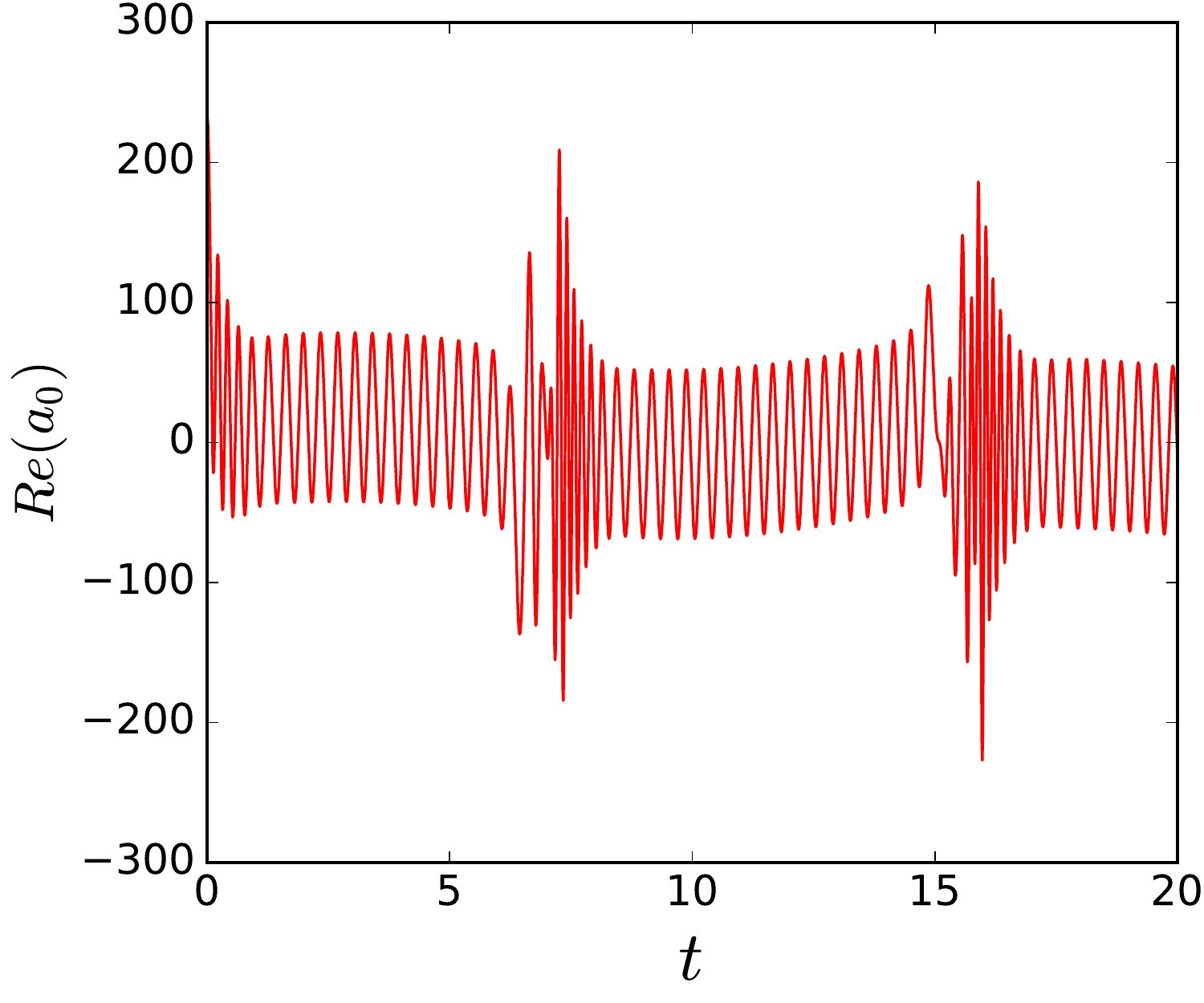}
  \caption{
    Real part of the zeroth Fourier mode of $A(x, t)$ versus time.
  }
  \label{fig:EIDCqcgl1d_phase_rotation}
\end{figure}

In \refSect{sect:exp1d}, we show that \ssa\ schemes slow down the integration of 
\fm\ parts, but the \sm\ parts still take the majority of computation time as we 
compare panel (b), (c) with panel (a) in \reffig{fig:EIDCqcgl1d_heat}. Also, 
\reffig{fig:EIDCqcgl1d_static_h}(b) shows that there is a plateau for the step 
sizes used in the \sm\ parts.   What prevents one from using a larger step size is the 
\emph{\fpr} of the complex field $A(x, t)$.
\refFig{fig:EIDCqcgl1d_phase_rotation} shows the real part of the zeroth Fourier 
mode $\re(a_0)$ of $A(x, t)$ during the integration period in
\reffig{fig:EIDCqcgl1d_heat}(a). The phase of $a_0$ rotates 
with a high frequency even though the profile $|A(x, t)|$ changes slowly in
the \sm\ parts. Thus, if \fpr\ of complex field $A(x, t)$ can be handled 
effectively, one can further accelerate the \sm\ parts.
We propose to use a \cmf\ which has a similar rotating frequency as the
original system to improve the results.   In particular, we show that 
the performance of ERK methods can be improved in this frame work. 

\subsection{Dynamics in a \cmf}  

To overcome the \fpr\ difficulty, we integrate the system in a comoving frame 
whose frequency is similar to the rotating frequency of $A(x, t)$.
Let
\begin{equation}
  \label{eq:comving_frame}
  A(x, t) = \Acm(x, t)e^{i\Omega t}
  \,,
\end{equation}
where $\Omega$ is the rotating frequency of this \cmf. $A(x, t)$ and $\Acm(x, t)$ is 
the state in the static and \cmf\ respectively. Substituting \eqref{eq:comving_frame}
into the one\dmn\ version of \eqref{eq:cqcgl}, we get
\begin{equation}
  \label{eq:cqcgl_comoving}
  \Acm_t  = (\mu - i\Omega) \Acm + (D_r + iD_i) \Acm_{xx}
  + (\beta_r + i\beta_i)|\Acm|^2\Acm
  + (\gamma_r + i\gamma_i)|\Acm|^4\Acm
  \,.
\end{equation}
By changing the real coefficient $\mu$ to the complex one $\mu - i\Omega$,  
we obtain the integrator in the \cmf.  Thus \cmf\ introduces nearly 
no additional computational cost compared with the integrator in the \stf. 
To find the frequency of this comoving frame, 
one can simply measure the rotating frequency of $A(x, t)$ in the whole 
domain $x\in[0, L]$ for a certain \sm\ part, 
then obtain an average rotating rate. However, this approach is hard to 
automate and other issues such as the \emph{phase-wrapping effect}, \ie, aliasing, can  
complicate this process.  

In this paper, we utilize the underlining dynamically invariant structure of this exploding 
phenomenon. 
\refFig{fig:EIDCqcgl1d_heat}(a) illustrates that  the basic structure of the 
dynamics is a \sm\ soliton which undergoes intermittent explosions.  If this
soliton is viewed as a traveling wave, then an exploding part can be regarded 
as one homoclinic orbit of this traveling wave.  We express an invariant solution of form
\begin{equation}
  \label{eq:traveling}
  A(x,t) = A_0(x+ct)e^{i\omega t}
  \,.
\end{equation}
Here, $A_0(x) = A(x, 0)$ is a localized field. Constants $c$ and $\omega$ are spatial 
translation and phase velocity respectively. Definition \eqref{eq:traveling} originates 
from the consideration of the two continuous symmetries of \cqcGLe, namely, equation 
\eqref{eq:cqcgl} is invariant under spatial translation $A(x,t) \to A(x+\ell,t)$ and 
phase rotation $A(x,t) \to e^{i\phi}A(x,t)$. Soliton explosions are the result of the rapid 
growth of perturbations in the unstable directions of a traveling wave. The collapse of 
explosions is due to the dispersion effect. For more descriptive details, 
see\rf{DingCvit16}. Therefore, we set the frequency $\Omega$ of the \cmf\ to the rotating 
frequency $\omega$ of this traveling wave, and integrate $\tilde{A}(x,t)$ instead of 
$A(x,t)$ for 
a better performance. 
 
\begin{figure}[h]
  \centering
  \includegraphics[width=0.95\textwidth]{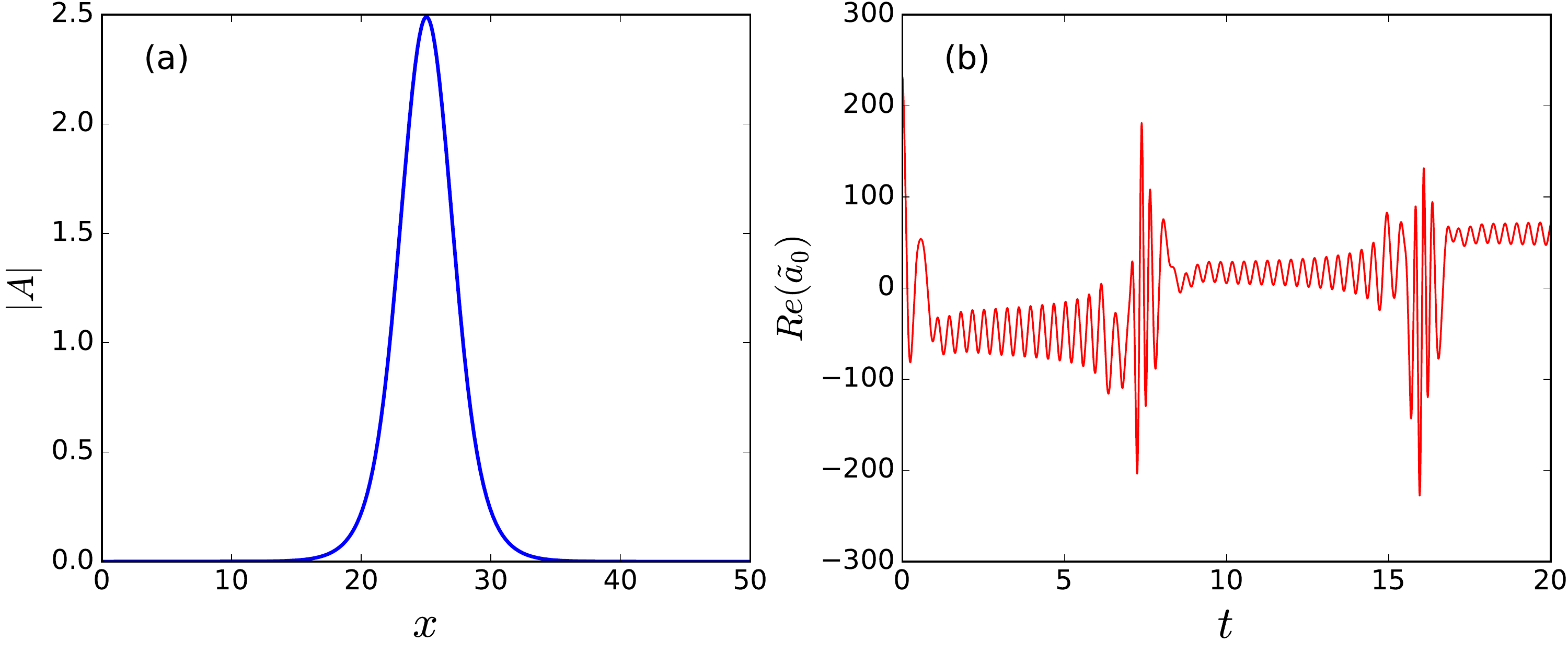}
  \caption{
    (a) Profile of the traveling wave.
    (b) Real part of the zeroth Fourier mode of $\tilde{A}(x, t)$ versus time.
  }
  \label{fig:EIDCqcgl1d_req_profile}
\end{figure}

We find traveling waves in the Fourier mode space, in which equation 
\eqref{eq:traveling} becomes 
\[
  a_k(t) = \exp(i\omega t + ikq_ct) \; a_k(0) \,,\quad
  k = 0, \pm 1, \pm 2, \dots 
  \,,
\]
where $q_c=2\pi c/L$.
Taking time derivative on both sides, we get
\begin{equation}
  \label{eq:req}
  \dot{a}_k(t) - ikq_c a_k -i \omega a_k = 0 \,,\quad
  k = 0, \pm 1, \pm 2, \dots 
  \,.
\end{equation}
Here velocity field $\dot{a}_k$ is given in \eqref{eq:cqcglFourier}. 
Equation \eqref{eq:req} defines an underdetermined system
with respect to variables $(a_k, c, \omega)$, whose roots are traveling waves. 
Given relatively good initial guesses, Newton-based methods converge quadratically to the 
traveling wave solution. In practice, we use Levenberg-Marquardt 
algorithm\rf{levenberg44, Marquardt63} for its good performance in solving underdetermined
systems. More details can be found in\rf{DingCvit16}. 
The traveling wave obtained
is shown in \reffig{fig:EIDCqcgl1d_req_profile}(a) with
\[
  c=0\,,\quad \omega=17.6675
  \,.
\]
This traveling wave lives in the symmetric subspace and thus has no spatial shift, 
but its phase rotates rapidly. By integrating in the \cmf\ $\Omega = \omega$, we obtain 
$\re(\acm_0)$ shown in \reffig{fig:EIDCqcgl1d_req_profile}(b). 
$\acm_0$ is the zeroth Fourier mode of $\Acm(x, t)$. Compared to 
\reffig{fig:EIDCqcgl1d_phase_rotation}, we see that the \fpr\ is effectively reduced 
for the \sm\ parts, while the explosion parts still have fast-phase dynamics. 

We emphasize that the traveling wave found for one specific set of parameters of 
the \cqcGLe\ can be used as an initial guess to find traveling waves in the parameter
space, so finding an appropriate frequency of the \cmf\ for different parameters 
can be easily automated. 

\subsection{Performance in the \cmf}

\begin{figure}[h]
  \centering
  \begin{minipage}{.3\textwidth}
    \centering \small{\texttt{(a)}}
    \includegraphics[width=\textwidth]{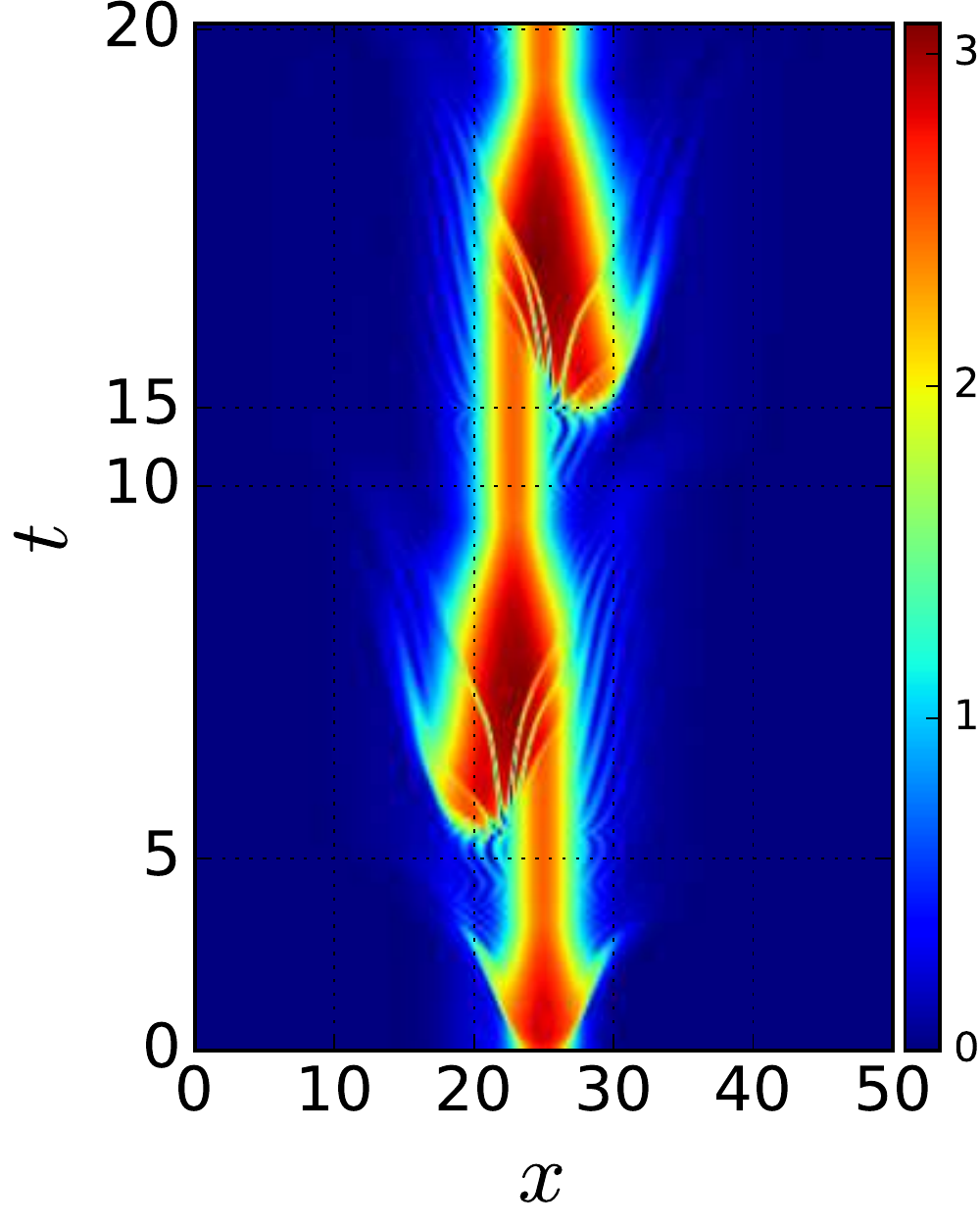}
  \end{minipage}
  \begin{minipage}{.45\textwidth}
    \centering \small{\texttt{(b)}}
    \includegraphics[width=\textwidth]{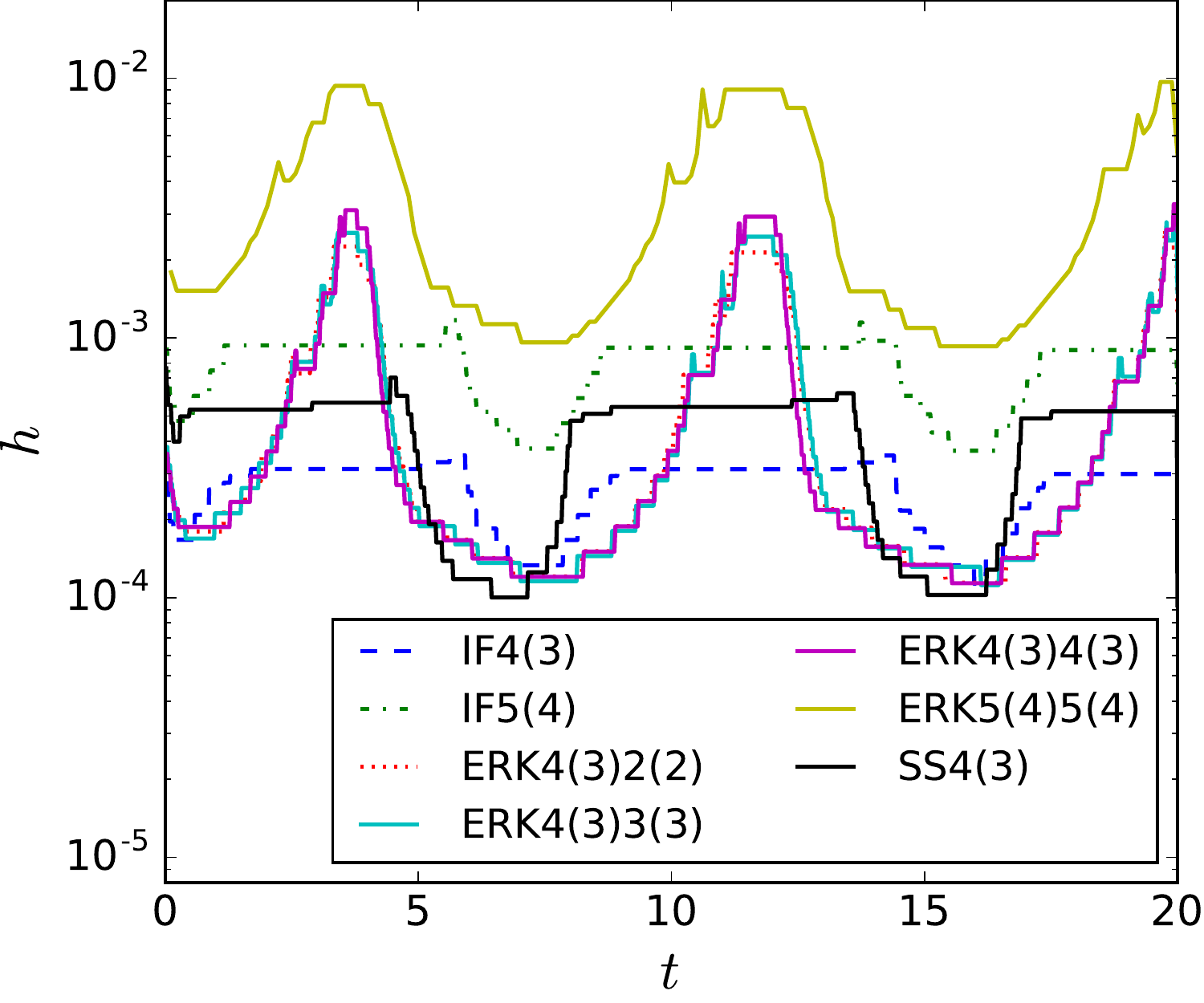}
  \end{minipage}
  \caption{
    (a) Spatiotemporal plot of $|A(x, t)|$ produced by ERK4(3)2(2) in the \cmf.
    (b) Step size $h$ vs $t$ with $\rtol=10^{-10}$ in the \cmf. 
  }
  \label{fig:EIDCcqcgl1d_hs_comoving}
\end{figure}

\refFig{fig:EIDCcqcgl1d_hs_comoving}(a) shows the integration result of ERK4(3)2(2) 
in the \cmf. Compared to \reffig{fig:EIDCqcgl1d_heat}(b)(c), 
the \sm\ part is  sufficiently accelerated.
Time steps used during integration process are shown in  
\reffig{fig:EIDCcqcgl1d_hs_comoving}(b). There are several interesting observations 
comparing \reffig{fig:EIDCcqcgl1d_hs_comoving}(b) with 
\reffig{fig:EIDCqcgl1d_static_h}(b).
First, for all seven methods, step sizes used during the \fm\ parts are almost the 
same $h\simeq 10^{-4}$ in both static and \cmf. This is reasonable because the \cmf\ 
cannot reduce the rapid phase rotation during explosions. 

Second, step sizes have increased substantially for ERK methods in the \cmf\ for the \sm\ 
parts, while there is no change for IF4(3)4, IF5(4) and SS4(3). The \cmf\ only 
promotes the performance of ERK methods, not IF or SS methods. This 
intuitively contradictory result comes from the difference between the intermediate 
states among different methods. The \cmf\ changes the linear part from $\mL$ to $\mL 
-i\Omega$ in \eqref{eq:cqcgl_comoving}. 
The intermediate state \eqref{eq:IF2} 
of an IF method
\[
 Y_i  = e^{hc_i\mL}y_n + h\sum_{j=1}^{i-1} a_{ij}e^{h\alpha_{ij}\mL}
  \mN(t_n + c_jh, Y_j)
\] 
and the two  Butcher tables, \reftab{tab:ifrk43} and \reftab{tab:ifrk54}, show that
coefficients $e^{hc_i\mL}$ and $a_{ij}e^{h\alpha_{ij}}$ 
shift by $e^{-i\Omega c_i h}$ and $e^{-i\Omega \alpha_{ij} h}$ respectively. 
Assume $Y_j$, $j < i$, has shift $e^{-i\Omega c_j h}$, then 
$a_{ij}e^{h\alpha_{ij}\mL} \mN(t_n + c_jh, Y_j)$
has shift $e^{-i\Omega c_i h}$ because $\alpha_{ij} + c_j = c_i$.  
The intermediate state at $t_n + c_ih$ changes from $Y_i$ to $Y_i e^{-i\Omega c_i h}$. 
Such a phase change only introduces a phase shift for the local error estimation in
\reftab{tab:lte}. 
Therefore, IF methods are invariant under $\mL \to \mL -i\Omega$. 
For SS methods, both \eqref{eq:SSy1} and \eqref{eq:SSy2} have the same phase 
shift, so transformation $\mL \to \mL -i\Omega$ only introduces a phase 
rotation for the local error estimation  and thus does not 
change the behavior of SS methods either. However, for ERK methods 
\eqref{eq:ETDRK}, coefficients $a_{ij}(h\mL)$ and $b_i(h\mL)$ are functions of
$\varphi_j(h\mL)$, which does not have an explicit phase rotation relation under 
transformation $\mL \to \mL -i\Omega$. So, intermediate state $Y_i$ is not transformed 
to $Ye^{-i\Omega c_i h}$. The \cmf\ modifies the local error estimation 
for ERK methods. For the \sm\ parts, the rapid phase rotation is effectively reduced 
in the \cmf\ and intermediate states tend to have smaller differences in phase. 
\refFig{fig:EIDCcqcgl1d_hs_comoving}(b) illustrates how  \cmf\ accelerates the 
integration of the \sm\ parts.

\begin{figure}[!ht]
  \centering
  \includegraphics[width=0.9\textwidth]{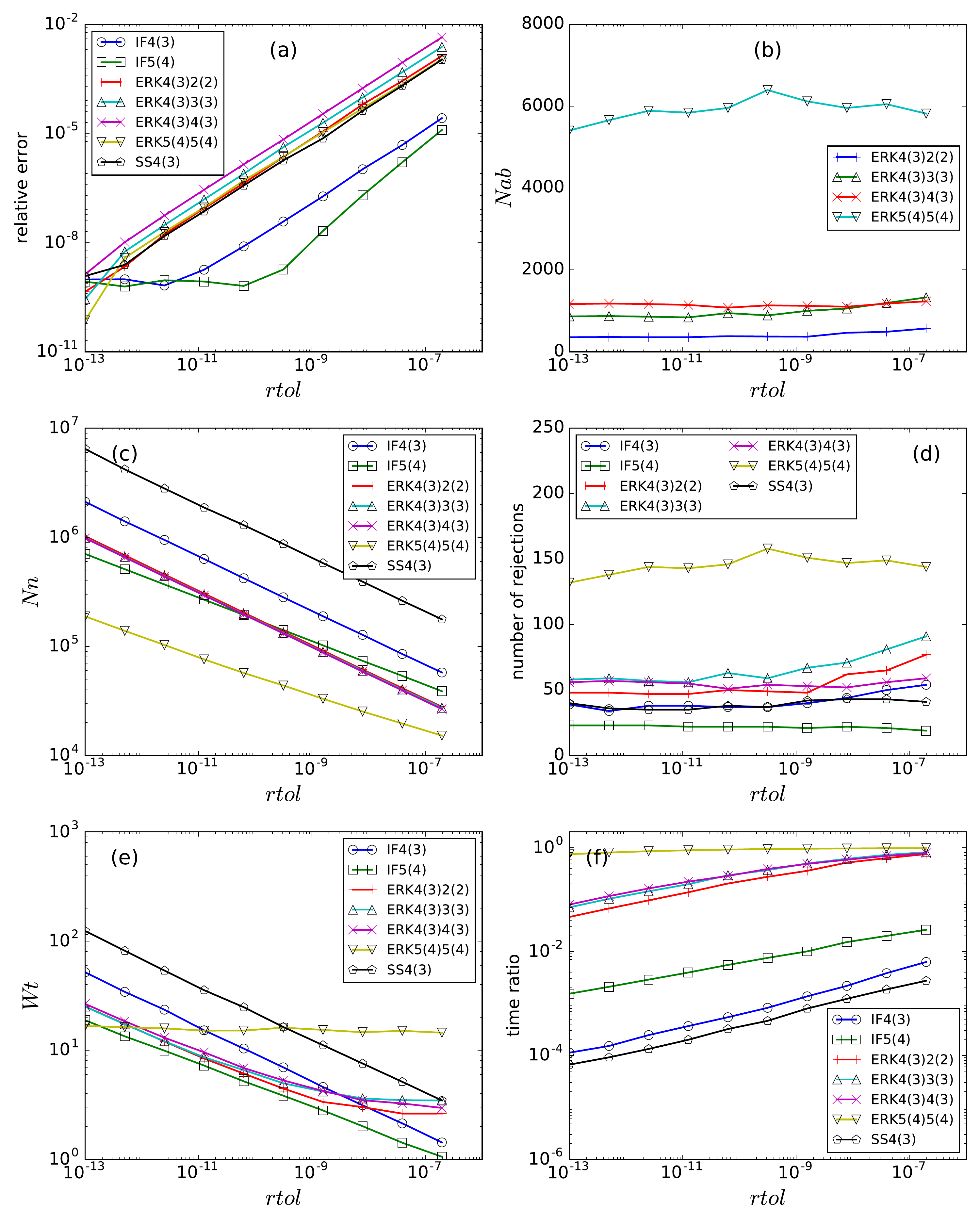}
  \caption{
    Same as \reffig{fig:EIDCqcgl1d_static} but in the \cmf.
  }
  \label{fig:EIDCqcgl1d_comoving}
\end{figure}

We repeat the numerical experiments of \reffig{fig:EIDCqcgl1d_static} in a \cmf\ shown in 
\refFig{fig:EIDCqcgl1d_comoving}. The performance 
of IF4(3), IF5(4) and SS4(3) does not change in the \cmf\ compared to that in the 
static frame. On the other hand, there are several differences for the ERK methods. 
Comparing \reffig{fig:EIDCqcgl1d_comoving}(a) with \reffig{fig:EIDCqcgl1d_static}(a), the 
global accuracy deteriorates in the \cmf\ since larger step sizes are used for the \sm\ 
parts. Smaller \rtol\ should be chosen in order to achieve the required global accuracy 
in the \cmf\ for ERK methods.  This is reasonable if one cares more about the percentage 
of time spent on the \fm\ parts. \refFig{fig:EIDCqcgl1d_comoving}(c) shows that the 
number of evaluation of $\mN(t, y)$ reduced significantly compared to the data in 
\reffig{fig:EIDCqcgl1d_static}(c). The total integration time decreased as shown in 
\reffig{fig:EIDCqcgl1d_comoving}(e). The number of times that a step size 
is rejected becomes 2 to 4 times larger in the \cmf\ as shown in 
\reffig{fig:EIDCqcgl1d_comoving}(d) compared to  \reffig{fig:EIDCqcgl1d_static}(d). 
For ERK5(4)5(4), the time for refilling the Butcher table takes the majority computation 
time even when \rtol\ is as small as $10^{-13}$, shown in 
\reffig{fig:EIDCqcgl1d_comoving}(f).  For a large \rtol, ERK5(4)5(4) is not as efficient 
as other schemes, as indicated by \reffig{fig:EIDCqcgl1d_comoving}(e).

In summary, the performance of ERK methods improves in a \cmf.  To spend more integration 
time on the \fm\ parts, one should use ERK4(3)2(2), ERK4(3)3(3), or ERK4(3)4(3), but not 
ERK5(4)5(4), since the last one spends too much time refilling its Butcher table.

\section{Two\dmn\ numerical experiments}
\label{sect:exp2d}
\begin{figure}[!ht]
  \centering
  \includegraphics[width=0.9\textwidth]{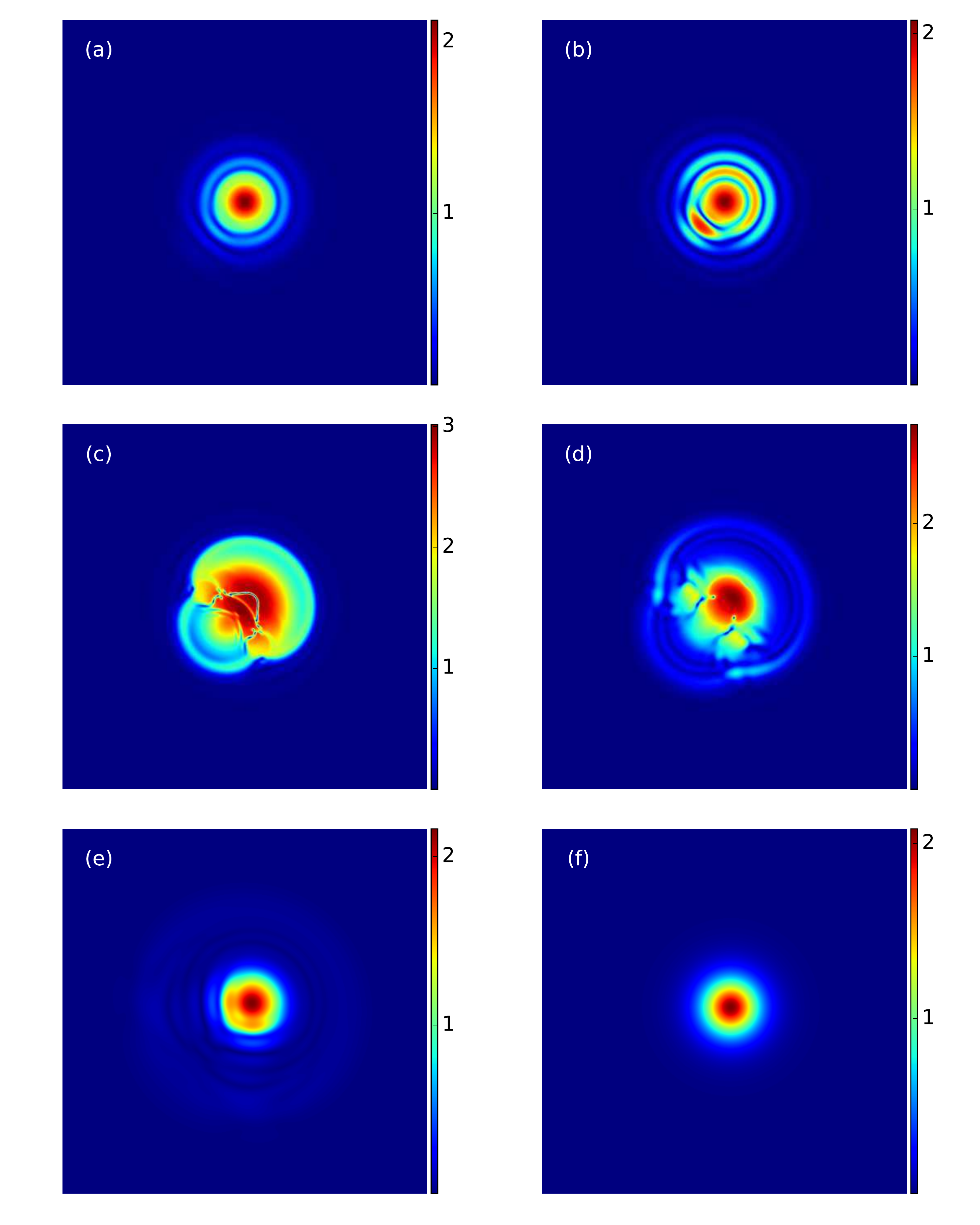}
  \caption{
    (a)$\sim$(e) Snapshots of one explosion in the two\dmn\ \cqcGLe\ using initial
    condition \eqref{eq:2dinit}. The system is integrated by ERK4(3)2(2). 
    These five snapshots correspond to the state at $t=3, 5, 8, 10, 12$ respectively.
    (f) The profile of an unstable traveling wave in this system.
    In all figures, the color represents the magnitude of $|A(x, y, t)|$. 
    System parameters are $\mu=-0.1$, $D_i=0.5$, $\gamma_r=-0.1$,
    $D_r=0.125$, $\beta_r=1$, $\beta_i=0.8$ and $\gamma_i=-0.6$.
  }
  \label{fig:EIDCqcgl2d_static_adapt_heat}
\end{figure}

\begin{figure}[!ht]
  \centering
  \includegraphics[width=0.9\textwidth]{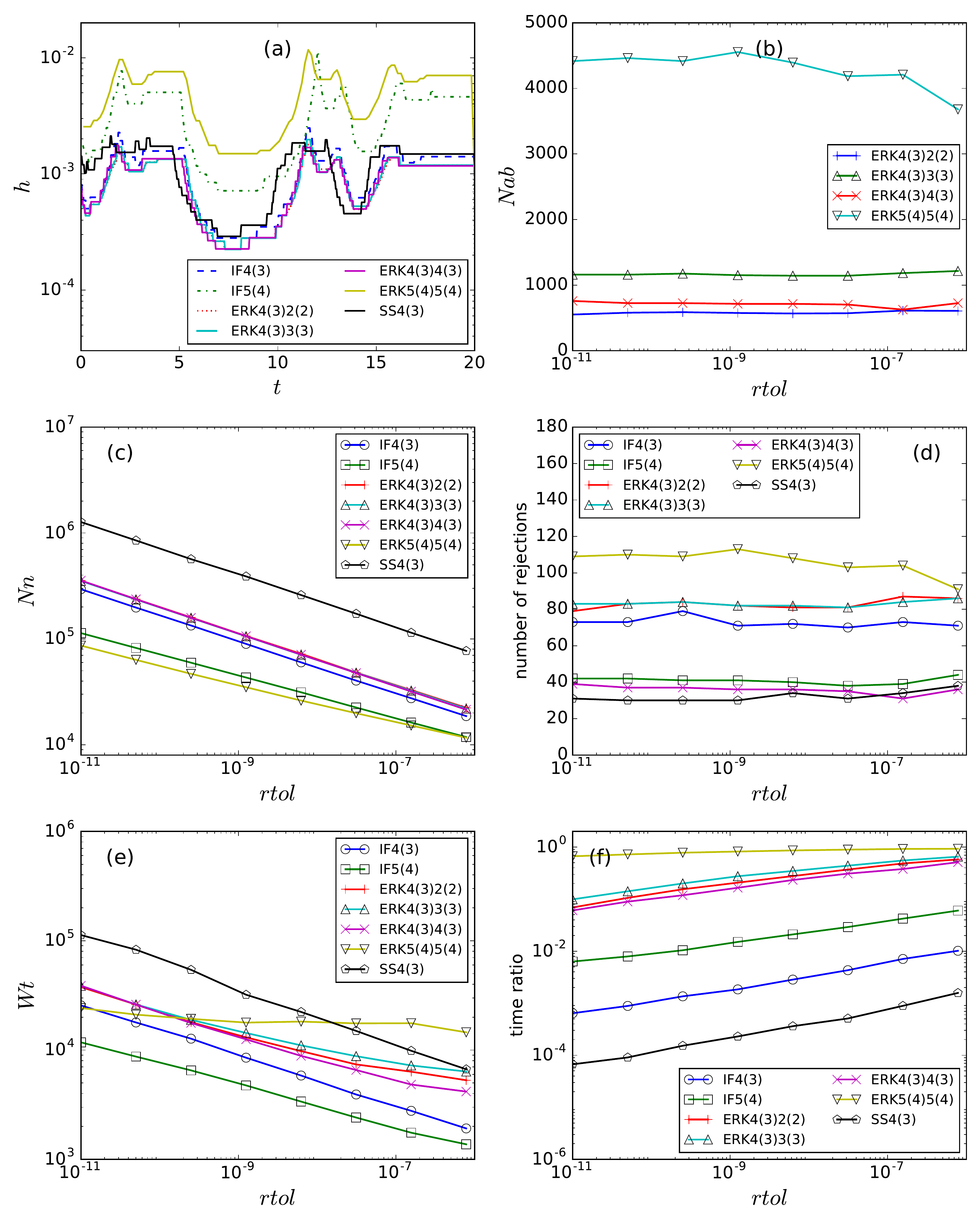}
  \caption{
    Performance of \ats\ schemes in the static frame for the two\dmn\ \cqcGLe.
    (a) Step sizes used during the integration process when $\rtol=10^{-9}$.
    (b)$\sim$(f) Performance measured by different metrics when $\rtol$ varies.
  }
  \label{fig:EIDCqcgl2d_static_statistic}
\end{figure}

\begin{figure}[!ht]
  \centering
  \includegraphics[width=0.9\textwidth]{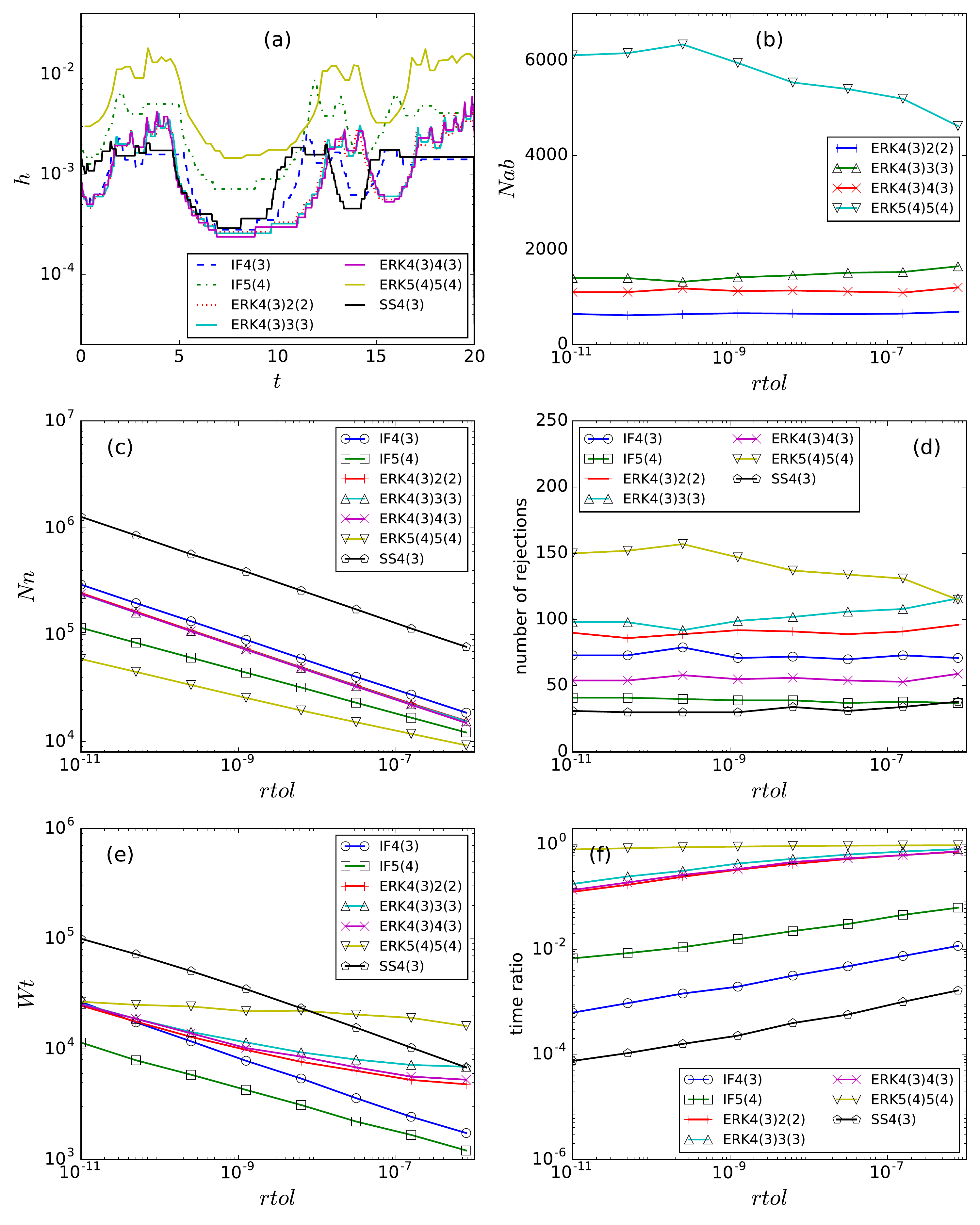}
  \caption{
    The same as \reffig{fig:EIDCqcgl2d_static_statistic} but in a comoving
    frame.
  }
  \label{fig:EIDCqcgl2d_comoving_statistic}
\end{figure}

In this section, we consider exploding solitons in two\dmn\ \cqcGLe.
All seven \ssa\ schemes are exploited and compared in both static and
\cmf.  
The following initial condition, which represents a Gaussian wave at the center of 
the grid with a small perturbation in the southwest direction, is used to
generate a sequence of exploding solitons.
\begin{equation}
  A(x, y, 0) = 2.5\exp\left(
    -450\left(\frac{x}{L} - \frac{1}{2} \right)^2 -450\left(\frac{y}{L} - \frac{1}{2} \right)^2
  \right) + 
  0.2\exp\left(
    -450\left(\frac{x}{L} - \frac{2}{5} \right)^2 -450\left(\frac{y}{L} - \frac{2}{5} \right)^2
  \right)
  \,.
  \label{eq:2dinit}
\end{equation}
We integrate the system for time window $t\in[0, 20]$, 
during which there are three explosions. The snapshots 
of one such explosion are shown in \reffig{fig:EIDCqcgl2d_static_adapt_heat} (a)$\sim$(e).
The profile of the soliton augments asymmetrically during this exploding process.
Also, there are several sharp cracks at the center of the soliton in panel (c), which
move and change from panel (c) to (d). All these fast-changing structures require a
smaller step size to maintain a certain local integration accuracy compared to the step
size needed before and after the explosion, \ie, panel (a) and (e) respectively. 

\refFig{fig:EIDCqcgl2d_static_statistic} shows the performance of the seven \ats\ 
schemes. Panel (a) shows the step sizes used during the integration. There are three dips 
in this figure, which represent that the integrators slow down during three exploding
instances. IF5(4) and ERK5(4)5(4) use larger step sizes than those of 4th-order schemes. 
All 4th-order schemes use similar step sizes. Panel (b)$\sim$(f) show the performance 
measured by different metrics. The tendencies shown in panel (b)$\sim$(f) are very 
similar to those in \reffig{fig:EIDCqcgl1d_static}. Among the four ERK methods, 
ERK5(4)5(4) has the largest number of times of calculating the elements of its Butcher 
table as shown in panel (b). Two facts account for this: First, there are 23 different 
elements in the Butcher table of ERK5(4)5(4), which are far more than other ERK methods. 
Second, as shown in panel (d), attempted trials of ERK5(4)5(4) are more likely to get 
rejected compare to other ERK methods.  Panel (c) shows the number of evaluations of the 
nonlinear function. Similar to \reffig{fig:EIDCqcgl1d_static}(c), the two 5th-order 
schemes have least evaluations of the nonlinear function. SS4(3) has the largest number 
of evaluations because it computes the nonlinear function 24 times in a single step. 
Panel (e) and (f) show the wall time of integration and the percentage of time spent on 
recalculating coefficients. The two\dmn\ simulation is about 3- or 4-order more expensive 
than the one\dmn\ simulation comparing panel (e) with \reffig{fig:EIDCqcgl1d_static}(e). 
In the two\dmn\ integration, ERK5(4)5(4) spends most of the time recalculating the 
coefficients. it is less efficient than IF5(4).

Similar to the one\dmn\ case, explosions in the two\dmn\
case can be visualized as homoclinic orbits of an unstable traveling wave of this
system.  A two\dmn\ traveling wave has form
\[
  A(x,y, t) = A_0(x+c_xt, y+c_yt)e^{i\omega t}
  \,,
\]
with $c_x$, $c_y$, and $\omega$ respectively the translation velocity in the $x$, $y$ 
direction, and the phase velocity. By using a shooting method\rf{ChaKer12}, we find 
a traveling wave living in the symmetric subspace with
\begin{equation}
  \label{eq:omega2d}
  c_x = 0\,, \quad c_y = 0\,, \quad \omega = 7.3982
  \,.  
\end{equation}
\refFig{fig:EIDCqcgl2d_static_adapt_heat}(f) shows the profile of this traveling wave.
For the \sm\ parts, the system has a similar phase-rotation rate as this traveling wave.
Therefore, similar to \eqref{eq:comving_frame} and \eqref{eq:cqcgl_comoving}, we can 
define a \cmf\ for the two\dmn\ \cqcGLe.
 
\refFig{fig:EIDCqcgl2d_comoving_statistic} displays the performance of the seven \ssa\
schemes in the \cmf. As the one\dmn\ case, only the performance of ERK schemes is
affected by the \cmf. Comparing panel (a) in \reffig{fig:EIDCqcgl2d_static_statistic} and
that in \reffig{fig:EIDCqcgl2d_comoving_statistic}, the step sizes used during the 
\sm\ parts are about 2 to 3 times larger for ERK schemes in the comoving frame, which
is manifest for the part after the third explosion. Also, for ERK schemes in panel  
(c), the number of times to evaluate the nonlinear function decreased insignificantly. 
However, the rejection rates increased for ERK schemes as shown in panel (d). As a 
consequence, the total integration time only decreased by a small amount as shown in 
panel (e). 
The improvement of performance in the \cmf\ is not striking. The reason is
that the phase velocity \eqref{eq:omega2d} is not large; thus the \fpr\ problem is not 
severe in the static frame. Therefore, to integrate the two\dmn\ \cqcGLe\ with the 
parameters chosen in this paper, the best choice out of the seven schemes is IF5(4).

\section{Conclusions}
\label{sect:concl}
We explored different \ssa\ schemes to integrate soliton solutions in
one- and two\dmn\ \cqcGLe. We put an emphasis on the exploding solitons which 
have different time scales. 
The \sm\ part and the \fm\ part should use different step sizes 
in order to integrate the system efficiently. 
By embedding lower order schemes in IF, ERK and SS methods, 
local integration error is controlled effectively.  The step size
is adapted to maintain a relative accuracy in each single integration step.
For solitons with extremely pulsating amplitudes, \tsa\ works well in the
static frame. While for exploding solitons,  
to better handle the \fpr\ difficulty, we integrated the system in a 
\cmf\ whose rotating frequency is similar to that of the soliton solution.
We show that integration in the \cmf\ can further accelerate the \sm\ parts 
for ERK methods. 

In the one\dmn\ case, the two 5th-order methods IF5(4) and ERK5(4)5(4) have 
the best performance. IF5(4) is easy to implement because it has a simple 
Butcher table structure. ERK5(4)5(4) can benefit from the comoving frame and 
remarkably slow down the \fm\ parts. Since ERK5(4)5(4) spends a long time 
refilling its Butcher table when local error tolerance \rtol\ becomes large,
we prefer ERK5(4)5(4) when \rtol\ is small but choose IF5(4) otherwise.
In the two\dmn\ case, we find that IF5(4) has the best performance. 
When the phase rotation is not very severe,  a \cmf\ may not improve the results much. 
When one need to implement a \ssa\ scheme quickly, then the 
4th-order schemes are suitable. 
ERK4(3)2(2) may be the best choice among all 4th-order methods considered in this paper 
because of its simple Butcher table structure. 
 
The main focus of this paper is to explore different methods to experiment \cqcGLe .
Nonetheless, these numerical schemes can be applied to other stiff (and non-stiff) 
systems that exhibit intermittent behaviors.

\section{Acknowledgements}
\label{sect:ack}

We are grateful to P. Cvitanovi\'c for providing insightful arguments about the
phase rotation phenomenon in one\dmn\ \cqcGLe.
X.Ding is supported by a grant from G. Robinson, Jr..
S.H. Kang is supported by Simons Foundation  grant 282311.

\appendix

\section{}
\label{sec:append}
\refTab{tab:5noc} displays the nonstiff 5th-order conditions.
There are 37 equations. The first 26 come from adding a black or white 
node to the root node of the 4th-order bi-colored trees\rf{Berland05}.
The remaining 11 equations are enumerated by counting the 2-fork type (8 equations),
the 3-fork type (2 equations) and the 4-fork type (1 equation).
For the general result to an arbitrary order, see \refref{Berland05} (Theorem 2.1).
\begin{table}[!h]
  \setlength\tabcolsep{3pt}
  \caption{Nonstiff 5th-order conditions.}
  \label{tab:5noc}
  \centering
  \begin{tabular}{|c | c | c|}
    \hline
    & Tree & condition \\
    \hline
    1 &
        \begin{tikzpicture}[mytree]
          \node [bn]{}    
          child {node[bn] {} 
            child {node[bn] {}}
            child {node[bn] {}} 
            child {node[bn] {}} 
          }; 
        \end{tikzpicture} 
           & $\sum \beta^{r, 0}\alpha_{r}^{j, 0}c_{j}^3 = \frac{1}{20}$ \\   
    2 & 
        \begin{tikzpicture}[mytree]
          \node [wn]{}    
          child {node[bn] {} 
            child {node[bn] {}}
            child {node[bn] {}} 
            child {node[bn] {}} 
          }; 
        \end{tikzpicture} 
           &   $\sum \beta^{r, 1}c_{r}^3 = \frac{1}{20}$  \\
    3 & 
        \begin{tikzpicture}[mytree]
          \node [bn]{}    
          child {node[bn] {}
            child {node[bn] {}}
            child {node[bn] {} 
              child {node[bn] {}}} 
          }; 
        \end{tikzpicture} 
           &   $\sum \beta^{r, 0}\alpha_{r}^{j,0}\alpha_{j}^{k,0}c_{j}c_{k} = \frac{1}{40}$  \\
    4 & 
        \begin{tikzpicture}[mytree]
          \node [wn]{}    
          child {node[bn] {}
            child {node[bn] {}}
            child {node[bn] {} 
              child {node[bn] {}}} 
          }; 
        \end{tikzpicture} 
           &   $\sum \beta^{r, 1}\alpha_{r}^{j,0}c_{r}c_{j} = \frac{1}{40}$  \\
    5 & 
        \begin{tikzpicture}[mytree]
          \node [bn]{}    
          child {node[bn] {}
            child {node[bn] {}}
            child {node[wn] {} 
              child {node[bn] {}}} 
          }; 
        \end{tikzpicture} 
           &   $\sum\beta^{r, 0}\alpha_{r}^{j,0}\alpha_{j}^{k,1}c_{j} = \frac{1}{40}$  \\
    6 & 
        \begin{tikzpicture}[mytree]
          \node [wn]{}    
          child {node[bn] {}
            child {node[bn] {}}
            child {node[wn] {} 
              child {node[bn] {}}} 
          }; 
        \end{tikzpicture} 
           &   $\sum \beta^{r, 1}\alpha_{r}^{j,1}c_{r} = \frac{1}{40}$  \\
    7 & 
        \begin{tikzpicture}[mytree]
          \node [bn]{}    
          child {node[bn] {}
            child {node[bn] {}
              child {node[bn] {}}
              child {node[bn] {}}} 
          }; 
        \end{tikzpicture} 
           &   $\sum \beta^{r, 0}\alpha_{r}^{j,0}\alpha_{j}^{k,0}c^2_{k} = \frac{1}{60}$  \\
    8 & 
        \begin{tikzpicture}[mytree]
          \node [wn]{}    
          child {node[bn] {}
            child {node[bn] {}
              child {node[bn] {}} 
              child {node[bn] {}}} 
          }; 
        \end{tikzpicture} 
           &   $\sum \beta^{r, 1}\alpha_{r}^{j,0}c^2_{j} = \frac{1}{60}$  \\
    9 & 
        \begin{tikzpicture}[mytree]
          \node [bn]{}    
          child {node[wn] {}
            child {node[bn] {}
              child {node[bn] {}} 
              child {node[bn] {}}} 
          }; 
        \end{tikzpicture} 
           &   $\sum \beta^{r, 0}\alpha_{r}^{j,1}c^2_{j} = \frac{1}{60}$  \\
    10 & 
         \begin{tikzpicture}[mytree]
           \node [wn]{}    
           child {node[wn] {}
             child {node[bn] {}
               child {node[bn] {}} 
               child {node[bn] {}}} 
           }; 
         \end{tikzpicture} 
           &   $\sum \beta^{r, 2}c^2_{r} = \frac{1}{60}$ \\
    11 & 
         \begin{tikzpicture}[mytree]
           \node [bn]{}    
           child {node[bn] {}
             child {node[bn] {}
               child {node[bn] {} 
                 child {node[bn] {}}}} 
           }; 
         \end{tikzpicture} 
           &   $\sum \beta^{r, 0}\alpha_{r}^{j,0}\alpha_{j}^{k,0}\alpha_{k}^{p,0}c_{p}
             = \frac{1}{120}$  \\
    12 & 
         \begin{tikzpicture}[mytree]
           \node [wn]{}    
           child {node[bn] {}
             child {node[bn] {}
               child {node[bn] {}
                 child {node[bn] {}}}} 
           }; 
         \end{tikzpicture} 
           &   $\sum \beta^{r, 1}\alpha_{r}^{j,0}\alpha_{j}^{k,0}c_{k} = \frac{1}{120}$ \\
    13 & 
         \begin{tikzpicture}[mytree]
           \node [bn]{}    
           child {node[bn] {}
             child {node[bn] {}
               child {node[wn] {} 
                 child {node[bn] {}}}} 
           }; 
         \end{tikzpicture} 
           &   $\sum \beta^{r, 0}\alpha_{r}^{j,0}\alpha_{j}^{k,0}\alpha_{k}^{p,1}
             = \frac{1}{120}$ \\ 
    14 & 
         \begin{tikzpicture}[mytree]
           \node [wn]{}    
           child {node[bn] {}
             child {node[bn] {}
               child {node[wn] {}
                 child {node[bn] {}}}} 
           }; 
         \end{tikzpicture} 
           &   $\sum \beta^{r, 1}\alpha_{r}^{j,0}\alpha_{j}^{k,1} = \frac{1}{120}$ \\
    \hline
  \end{tabular}
  \begin{tabular}{|c | c | c|}
    \hline
    & Tree & condition \\
    \hline
    15 & 
         \begin{tikzpicture}[mytree]
           \node [bn]{}    
           child {node[bn] {}
             child {node[wn] {}
               child {node[bn] {} 
                 child {node[bn] {}}}} 
           }; 
         \end{tikzpicture} 
           &   $\sum \beta^{r, 0}\alpha_{r}^{j,0}\alpha_{j}^{k,1}c_k = \frac{1}{120}$  \\
    16 & 
         \begin{tikzpicture}[mytree]
           \node [wn]{}    
           child {node[bn] {}
             child {node[wn] {}
               child {node[bn] {}
                 child {node[bn] {}}}} 
           }; 
         \end{tikzpicture} 
           &   $\sum \beta^{r, 1}\alpha_{r}^{j,1}c_{j} = \frac{1}{120}$ \\
    17 & 
         \begin{tikzpicture}[mytree]
           \node [bn]{}    
           child {node[bn] {}
             child {node[wn] {}
               child {node[wn] {} 
                 child {node[bn] {}}}} 
           }; 
         \end{tikzpicture} 
           &   $\sum \beta^{r, 0}\alpha_{r}^{j,0}\alpha_{j}^{k,2} = \frac{1}{120}$  \\
    18 & 
         \begin{tikzpicture}[mytree]
           \node [wn]{}    
           child {node[bn] {}
             child {node[wn] {}
               child {node[wn] {}
                 child {node[bn] {}}}} 
           }; 
         \end{tikzpicture} 
           &   $\sum \beta^{r, 1}\alpha_{r}^{j,2} = \frac{1}{120}$ \\
    19 & 
         \begin{tikzpicture}[mytree]
           \node [bn]{}    
           child {node[wn] {}
             child {node[bn] {}
               child {node[bn] {} 
                 child {node[bn] {}}}} 
           }; 
         \end{tikzpicture} 
           &   $\sum \beta^{r, 0}\alpha_{r}^{j,1}\alpha_{j}^{k,0}c_{k} = \frac{1}{120}$  \\
    20 & 
         \begin{tikzpicture}[mytree]
           \node [wn]{}    
           child {node[wn] {}
             child {node[bn] {}
               child {node[bn] {}
                 child {node[bn] {}}}} 
           }; 
         \end{tikzpicture} 
           &   $\sum \beta^{r, 2}\alpha_{r}^{j,0}c_j = \frac{1}{120}$ \\
    21 & 
         \begin{tikzpicture}[mytree]
           \node [bn]{}    
           child {node[wn] {}
             child {node[bn] {}
               child {node[wn] {} 
                 child {node[bn] {}}}} 
           }; 
         \end{tikzpicture} 
           &   $\sum \beta^{r, 0}\alpha_{r}^{j,1}\alpha_{j}^{k,1} = \frac{1}{120}$  \\
    22 & 
         \begin{tikzpicture}[mytree]
           \node [wn]{}    
           child {node[wn] {}
             child {node[bn] {}
               child {node[wn] {}
                 child {node[bn] {}}}} 
           }; 
         \end{tikzpicture} 
           &   $\sum \beta^{r, 2}\alpha_{r}^{j,1} = \frac{1}{120}$ \\
    23 & 
         \begin{tikzpicture}[mytree]
           \node [bn]{}    
           child {node[wn] {}
             child {node[wn] {}
               child {node[bn] {} 
                 child {node[bn] {}}}} 
           }; 
         \end{tikzpicture} 
           &   $\sum \beta^{r, 0}\alpha_{r}^{j,2}c_j = \frac{1}{120}$  \\
    24 & 
         \begin{tikzpicture}[mytree]
           \node [wn]{}    
           child {node[wn] {}
             child {node[wn] {}
               child {node[bn] {}
                 child {node[bn] {}}}} 
           }; 
         \end{tikzpicture} 
           &   $\sum \beta^{r, 3}c_r = \frac{1}{120}$ \\
    25 & 
         \begin{tikzpicture}[mytree]
           \node [bn]{}    
           child {node[wn] {}
             child {node[wn] {}
               child {node[wn] {}
                 child {node[bn] {}}}} 
           }; 
         \end{tikzpicture} 
           &   $\sum \beta^{r, 0}\alpha_{r}^{j,3} = \frac{1}{120}$ \\
    26 & 
         \begin{tikzpicture}[mytree]
           \node [wn]{}    
           child {node[wn] {}
             child {node[wn] {}
               child {node[wn] {} 
                 child {node[bn] {}}}} 
           }; 
         \end{tikzpicture} 
           &   $\sum \beta^{r, 4} = \frac{1}{120}$  \\
    \hline
  \end{tabular}   
  \begin{tabular}{|c | c | c|}
    \hline
    & Tree & condition \\
    \hline
    27 & 
         \begin{tikzpicture}[mytree]
           \node [bn]{}    
           child {node[bn] {}}
           child {node[bn] {}
             child {node[bn] {}}
             child {node[bn] {}}
           }; 
         \end{tikzpicture} 
           &   $\sum \beta^{r, 0}\alpha_{r}^{j,0}c_j^2c_r = \frac{1}{15}$ \\
    28 & 
         \begin{tikzpicture}[mytree]
           \node [bn]{}    
           child {node[bn] {}}
           child {node[bn] {}
             child {node[bn] {} 
               child {node[bn] {}}}
           }; 
         \end{tikzpicture} 
           &   $\sum \beta^{r, 0}\alpha_{r}^{j,0}\alpha_{j}^{k,0}c_kc_r = \frac{1}{30}$  \\
    29 & 
         \begin{tikzpicture}[mytree]
           \node [bn]{}    
           child {node[bn] {}}
           child {node[bn] {}
             child {node[wn] {}
               child {node[bn] {}}}
           }; 
         \end{tikzpicture} 
           &   $\sum \beta^{r, 0}\alpha_{r}^{j,0}\alpha_{j}^{k,1}c_r = \frac{1}{30}$ \\
    30 & 
         \begin{tikzpicture}[mytree]
           \node [bn]{}   
           child {node[bn] {}
             child {node[bn] {}}}
           child {node[bn] {}
             child {node[bn]{}} 
           }; 
         \end{tikzpicture} 
           &   $\sum \beta^{r, 0}\alpha_{r}^{j,0}c_j \alpha_{r}^{k,0}c_k = \frac{1}{20}$  \\
    31 & 
         \begin{tikzpicture}[mytree]
           \node [bn]{}    
           child {node[wn] {}
             child {node[bn] {}}}
           child {node[wn] {}
             child {node[bn]{}} 
           }; 
         \end{tikzpicture} 
           &   $\sum \beta^{r, 0}\alpha_{r}^{j,1}\alpha_{r}^{k,1} = \frac{1}{20}$ \\
    32 & 
         \begin{tikzpicture}[mytree]
           \node [bn]{}   
           child {node[wn] {}
             child {node[bn] {}}}
           child {node[bn] {}
             child {node[bn]{}} 
           }; 
         \end{tikzpicture} 
           &   $\sum \beta^{r, 0}\alpha_{r}^{j,0}c_j \alpha_{r}^{k,1}= \frac{1}{20}$  \\
    33 & 
         \begin{tikzpicture}[mytree]
           \node [bn]{}    
           child {node[bn] {}}
           child {node[wn] {}
             child {node[bn] {} 
               child {node[bn] {}}}
           }; 
         \end{tikzpicture} 
           &   $\sum \beta^{r, 0}c_r\alpha_{r}^{j,1}c_j = \frac{1}{30}$ \\
    34 & 
         \begin{tikzpicture}[mytree]
           \node [bn]{}    
           child {node[bn] {}}
           child {node[wn] {}
             child {node[wn] {} 
               child {node[bn] {}}}
           }; 
         \end{tikzpicture} 
           &   $\sum \beta^{r, 0}c_r\alpha_{r}^{j,2} = \frac{1}{30}$  \\
    35 & 
         \begin{tikzpicture}[mytree]
           \node [bn]{}   
           child {node[bn] {}}
           child {node[bn] {}}
           child {node[bn] {} 
             child {node[bn] {}}
           }; 
         \end{tikzpicture} 
           &   $\sum \beta^{r, 0}c_r^2\alpha_{r}^{j,0}c_j = \frac{1}{10}$  \\
    36 & 
         \begin{tikzpicture}[mytree]
           \node [bn]{}   
           child {node[bn] {}}
           child {node[bn] {}}
           child {node[wn] {} 
             child {node[bn] {}}
           }; 
         \end{tikzpicture} 
           &  $\sum \beta^{r, 0}c_r^2\alpha_{r}^{j,1} = \frac{1}{10}$  \\
    37 & 
         \begin{tikzpicture}[mytree]
           \node [bn]{}   
           child {node[bn] {}}
           child {node[bn] {}}
           child {node[bn] {}}
           child {node[bn] {}
           }; 
         \end{tikzpicture} 
           &    $\sum \beta^{r, 0}c_r^4 = \frac{1}{5}$  \\
    \hline
  \end{tabular}
\end{table}

\section*{References}
\bibliographystyle{elsarticle-num}
\bibliography{DingKang16.bbl}


\end{document}